\documentclass[12pt]{JHEP3}
\usepackage{amsmath}

\newcommand{\e}{\epsilon}

\renewcommand{\L}{{\mathcal{L}}}
\newcommand{\bL}{\bar{{\mathcal{L}}}}

\renewcommand\O{{\mathcal{O}}}

\newcommand{\be}[1]{ \begin{equation}\label{#1} }
\newcommand{\ee}{\end{equation}}
\newcommand{\ben}[1]{\begin{eqnarray}\label{#1} }
\newcommand{\een}{\end{eqnarray}}
\newcommand{\eq}[1]{(\ref{#1})}
\def\ZZZ{{\hskip-3pt\hbox{ Z\kern-1.6mm Z}}}
\def\zzz{{\hskip-3pt\hbox{ z\kern-1mm z}}}

\newcommand{\p}{\partial}

\newcommand{\D}{\Delta}

\newcommand{\refb}[1]{(\ref{#1})}

\def\one{{\hbox{ 1\kern-.8mm l}}}
\def\zero{{\hbox{ 0\kern-1.5mm 0}}}

\title{Supersymmetric Extension of GCA in 2d}

\author{
Ipsita Mandal$^{1,2}$\\
$\;$ $^1$Harish-Chandra Research Institute,\\
$\;$ $\,$Chhatnag Road, Jhusi,\\
$\;$ $\,$Allahabad 211019, India\\

$\;$ $^2$LPTHE, Universite Pierre et Marie Curie,\\
$\;$ $\,$Paris 6, 4 Place Jussieu,\\
$\;$ $\,$75252 Paris Cedex 05, France\\

$\;$\email{ipsita@hri.res.in}
}

\abstract{We derive the infinite dimensional Supersymmetric Galilean Conformal Algebra (SGCA) in the case of two spacetime dimensions by performing group contraction on 2d superconformal algebra. We also obtain the representations of the generators in terms of superspace coordinates. Here we find realisations of the SGCA by considering scaling limits of certain 2d SCFTs which are non-unitary and have their left and right central charges become large in magnitude and opposite in sign. We focus on the Neveu-Schwarz sector of the parent SCFTs and develop, in parallel to the GCA studies recently in (hep-th/0912.1090), the representation theory based on SGCA primaries, Ward identities for their correlation functions and their descendants which are null states.}

\preprint{HRI/ST/1004}

\begin{document}

\baselineskip 3.5ex

\section{Introduction}

The study of conformal field theories has continued to be a major field of research ever since the AdS/CFT correspondence was realised as a potentially powerful tool in studying real-life systems. Mainly these studies have been based on the theories having conformal symmetry in relativistic spacetime. Non relativistic versions of conformal symmetry have been somewhat studied in the context of the so-called Schrodinger symmetry \cite{Hagen:1972pd, Niederer:1972zz, Henkel:1993sg, Nishida:2007pj} , which is an enhanced symmetry that arises in taking the nonrelativistic limit of the massive Klein-Gordon equation. In $d$ spacetime dimensions, this group has, in addition to the ${d(d+1)\over 2}$ parameters of the Galilean group, a nonrelativistic dilatation $x_i \rightarrow \lambda x_i, t\rightarrow \lambda^2 t$ and {\it one} more conformal generator. Recently, there has been considerable work involving nonrelativistic limits of the full $SO(d,2)$ conformal invariance. In fact, a non-relativistic group contraction of $SO(d,2)$ gives rise to a ${(d+1)(d+2)\over 2}$ parameter group (see, e.g., \cite{negro1997, Lukierski:2005xy} and in an $AdS$ context \cite{Gomis:2005pg}). This contains the Galilean symmetries together with a uniform dilatation  $x_i \rightarrow \lambda x_i\,, \,t\rightarrow \lambda t$ and  $d$ other generators which are the analogues of the special conformal transformations. 

It has been shown in \cite{Bagchi:2009my} that the actual set of conformal isometries of nonrelativistic spacetime is much larger than the above contracted version of $SO(d,2)$ . In {\it every} spacetime dimension it is actually an infinite dimensional algebra which was dubbed the Galilean Conformal Algebra (GCA). This algebra consists of local conformal transformations acting on time (generating a single copy of the Virasoro algebra), together with a current algebra for rotations as well as arbitrary time-dependent boosts in the spatial directions. A mathematically identical infinite dimensional algebra (called ${\rm altv}_1$) in the case of one spatial dimension had also appeared independently in \cite{Henkel:2002vd} in the context of statistical mechanics. It was also noted in \cite{Henkel06}
that $altv_1$ arises mathematically as a contraction of (two copies of) the  Virasoro algebra. It was shown in \cite{Duval:2009vt} that the infinite dimensional algebra in \cite{Bagchi:2009my} can also be obtained by considering the natural nonrelativistic limit of the relativistic conformal killing equations and is the maximal set of such nonrelativistic conformal isometries.\footnote{For related work on various aspects of the GCA, see \cite{Fouxon:2008tb}--\cite{IPM}.} The situation is very analogous to that in the relativistic $d=2$ case, where one has two copies of the Virasoro algebra generating the maximal set of local conformal isometries.

The analysis in \cite{Bagchi:2009my} was entirely classical, whereas in \cite{Bagchi:2009ca} (see also \cite{Alishahiha:2009np, Martelli:2009uc}) the two and three point correlation functions (of
primary fields) were obtained as solutions of the Ward identities for
the finite part of the GCA (which arises as the contraction of $SO(d,2)$). Finally in \cite{2dgca}, the quantum mechanical realisation of the GCA in two dimensions was studied in great detail, where 2d GCFTs with nonzero central charges were obtained by considering a somewhat unusual limit of non-unitary 2d CFTs.

Recently, supersymmetric extension of GCA in 4d was considered in \cite{Bagchi:2009ke} and \cite{deAzcarraga:2009ch}. In \cite{Sakaguchi:2009de}, SuperGCA in 3, 4 and 6 dimensions was studied. In the present paper, we study the $N=(1,1)$ supersymmetric extension of GCFTs in 2d, dubbed ``SGCFT''. Most of the algebraic structures of the 2d CFTs can be extended to their supersymmetric extensions, and the associated representation theory can also be developed along similar lines. The superconformal symmetries are also relevant for the superstring theory and the Tricritical Ising Model. We refer the reader to \cite{fqs1984}-\cite{gko} (and references therein) for an extensive study of the 2d superconformal theories. 

As in \cite{2dgca}, the families of 2d SCFTs we will need to consider are rather unusual in that their left and right central charges, $c$ and $\bar c$, are scaled (as we take the nonrelativistic limit) such that their magnitudes go to infinity but are opposite in sign. The parent theories are thus necessarily non-unitary and, not unsurprisingly, this non-unitariness is inherited by the daughter GCFTs. Since non-unitary 2d CFTs arise in a number of contexts in statistical mechanics as well as string theory, one might expect that the 2d GCFTs realised here would also be interesting objects to study.

In the present work, we focus our attention on the Neveu-Schwarz (NS) sector. Our study of 2d SGCAs in this paper proceeds along two parallel lines. The first line of development is as  described above and consists of taking carefully the nonrelativistic scaling limit of the parent 2d SCFT. We find that this limit, while unusual, appears to give sensible answers. Specifically, we will study in this way, the representation theory (including null vectors), the Ward identities, fusion rules, and finally the equations for correlation functions following from the existence of level $\frac{3}{2}$ null states. In all these cases we find that a non-trivial scaling limit of the 2d SCFTs exists. This is not {\it a priori} obvious, since the limit involves keeping terms both of ${\cal O}({1\over \e})$ and of ${\cal O}(1)$ (where $\e$ is the scaling parameter which is taken to zero). The second line of development obtains many of these same results by carrying out an autonomous analysis of the SGCA, i.e., independent of the above limiting procedure.

The structure of the paper is as follows: In the next section, we discuss how the generators of the 2d SGCA arise from a group contraction of (combinations of) the usual holomorphic and anti-holomorphic superconformal sectors. In Sec.~3, we proceed to construct representations of the 2d SGCA in a manner analogous to the NS sector of the SuperVirasoro representation theory, defining primaries and descendants. The primaries are labelled by a conformal weight $\Delta$ and a boost eigenvalue $\xi$. We also show that the state space is generically non-unitary. Sec.~4 deals with the non-relativistic Ward identities focussing on the case of two and three point functions.

We go back to the representation theory in Sec.~5 to consider null vectors of the SGCA. We explicitly find the conditions for having null states at level $\frac{3}{2}$ and check the corresponding conditions derivable from the scaling limit of the SuperVirasoro algebra. We also take the scaling limit of the Kac table for null states at arbitrary level. From Sec.~6 onwards, we focus on SGCA primaries taking values in the nonrelativistic Kac table. We derive the general differential equations for an $n$-point correlator which follow from the existence of level $\frac{3}{2}$ null states in Sec.~6. Finally in Sec.~7,  using the differential equations of Sec.~6, we proceed to derive the SGCA fusion rules that follow from the SGCA three point functions. Appendix A discusses some issues regarding SGCA descendants and their conformal blocks.

\section{2d SGCA from Group Contraction}
In this section, we derive the supersymmetric extension of the GCA in 2d, by performing group contraction on the 2d superconformal algebra studied in \cite{fqs1984}-\cite{gko}.

\subsection{Review of GCA in Arbitrary Dimensions}
The maximal set of conformal isometries of Galilean spacetime generates the infinite dimensional Galilean Conformal Algebra \cite{Bagchi:2009my}. The notion of Galilean spacetime is a little subtle since the spacetime metric degenerates into a spatial part and a temporal part. Nevertheless there is a definite limiting sense (of the relativistic spacetime) in which one can define the conformal isometries (see \cite{Duval:2009vt}) of the nonrelativistic geometry. Algebraically, the set of vector fields generating these symmetries 
are given by:

\ben{gcavec}
L^{(n)} &=& -(n+1)t^nx_i\p_i -t^{n+1}\p_t \,,\cr
M_i^{(n)} &=& t^{n+1}\p_i\,, \cr
J_a^{(n)} \equiv J_{ij}^{(n)} &= & -t^n(x_i\p_j-x_j\p_i)\,,
\een 
for $n \in \mathcal{Z}$. Here $i=1,\,2,\ldots,\, (d-1)$ range over the spatial directions. 
These vector fields obey the algebra:
\ben{vkmalg}
[L^{(m)}, L^{(n)}] &=& (m-n)L^{(m+n)}, \qquad [L^{(m)}, J_{a}^{(n)}] = -n J_{a}^{(m+n)}, \cr
[J_a^{(n)}, J_b^{(m)}]&=& f_{abc}J_c^{(n+m)}, \qquad  [L^{(m)}, M_i^{(n)}] =(m-n)M_i^{(m+n)}. 
\een

\subsection{SGCA from SuperVirasoro in 2d}

The finite dimensional subalgebra of the GCA (also sometimes
referred to as the GCA), which consists of taking $n=0,\pm1$ for the
$L^{(n)}, M_i^{(n)}$ together with $J_a^{(0)}$, is obtained
by considering the nonrelativistic  contraction of the usual (finite
dimensional) global conformal algebra $SO(d,2)$ (in $d>2$ spacetime
dimensions) 
(see for example \cite{negro1997}--\cite{Bagchi:2009my}). 

However, in two spacetime dimensions, as is well known, the situation is special. The relativistic conformal algebra is infinite dimensional and consists of two copies of the Virasoro algebra. In \cite{2dgca}, GCA with central charges was realised by taking a special limit of a non-unitary relativistic 2d CFT.

Here we take a similar limit on the 2d relativistic superconformal algebra, which is also infinite dimensional and consists of two copies of the SuperVirasoro algebra.

The two copies of the SuperVirasoro algebra are given by:
\ben{relsalg}
[\L_m, \L_n] &=& (m-n) \L_{m+n} + {c \over 8} m(m^2-1) \delta_{m+n,0}\,,\crcr
[\L_m, {\mathcal{G}}_r] &=& (\frac{1}{2} m - r) {\mathcal{G}}_{m+r}\,\crcr
\lbrace {\mathcal{G}}_r , {\mathcal{G}}_s \rbrace &=& 2 \L_{r+s} + {c \over 2} (r^2- \frac{1}{4}) \delta_{r+s,0}\,,\crcr
[\bL_m, \bL_n] &=& (m-n) \bL_{m+n} + {\bar c \over 8} m(m^2-1)\delta_{m+n,0}\,,\crcr
[ {\bar{\mathcal{L}}}_m, {\bar{\mathcal{G}}}_r ] &=& (\frac{1}{2} m - r) {\bar{\mathcal{G}}}_{m+r}\,,\crcr
\lbrace {\bar{\mathcal{G}}}_r , {\bar{\mathcal{G}}}_s\rbrace &=& 2 {\bar{\mathcal{L}}}_{r+s} + {\bar c \over 2} (r^2- \frac{1}{4}) \delta_{r+s,0}\,, 
\een 
where $m,n\,\in \mathcal{Z}$ and \textit{either} $r,s \,\in \, \mathcal{Z}$ [Ramond case] \textit{or} $r,s\, \in \, \mathcal{Z}+\frac{1}{2}$ [Neveu-Schwarz case].

We now perform group contraction with the new generators defined as:
\ben{lincom}
L_n &=& \lim_{\epsilon \to 0}\, ( \bar{\mathcal{L}}_n + \mathcal{L}_n)\,, \quad M_n = \lim_{\epsilon \to 0}\, \epsilon \,( \bar{\mathcal{L}}_n - \mathcal{L}_n)\,, \crcr
G_n &=& \lim_{\epsilon \to 0}\, ( {\bar{\mathcal{G}}}_n + {\mathcal{G}}_n) \,, \quad H_n = \lim_{\epsilon \to 0}\, \epsilon \,( {\bar{\mathcal{G}}}_n  - {\mathcal{G}}_n) \,,
\een
where for the bosonic part we have followed \cite{2dgca}, and for the fermionic part we have chosen a limit so as to get all the bosonic generators as anticommutators of the fermionic ones (here we have followed the scaling used in \cite{Sakaguchi:2009de}).

The above generators define the SGCA and obey the algebra:

\ben{sgca}
[L_m , L_n] &=&  (m-n) L_{m+n} + C_1 m (m^2-1) \delta_{m+n,0}\,,\crcr
[L_{m}, M_{n}] &=& (m-n) M_{m+n} + C_2 m(m^2-1) \delta_{m+n,0}\,, \crcr 
[M_{m}, M_{n}] &=& 0 \,, \crcr
\lbrace G_r, G_s\rbrace &=&  2 L_{r+s} + 4 C_1 (r^2- \frac{1}{4}) \delta_{r+s,0}\,,\crcr
\lbrace H_r , H_s \rbrace &=& 0 \,,\crcr
\lbrace G_r,H_s \rbrace &=& 2 M_{r+s} + 4 C_2 (r^2- \frac{1}{4}) \delta_{r+s,0}\,,\crcr
[L_m ,G_r] &=&  (\frac{1}{2} m - r) G_{m+r} \,,\quad [L_m ,H_r] =  (\frac{1}{2} m - r) H_{m+r} \,, \crcr
[M_m ,G_r] &=&  (\frac{1}{2} m - r) H_{m+r} \,,\quad [M_m ,H_r] =0 \,,
\een{}
where the central charges are given by:
\be{centch}
C_1 = \lim_{\epsilon \to 0} {{\bar c + c} \over 8}\,, \quad C_2 = \lim_{\epsilon \to 0} \e\,{{\bar c-c} \over 8}\,.
\ee
Thus, for a non-zero $C_2$ in the limit $\e\rightarrow 0$, we see that we need $\bar c-c \propto
\O({1\over \e})$. At the same time, requiring $C_1$ to be finite, we find
that  $c+\bar c$ should be $\O(1)$. 
As in \cite{2dgca}, we will make the slightly stronger assumption that
$\bar c - c = {\cal O}({1 \over \epsilon}) + {\cal O}(\epsilon)$.) Actually this is motivated by the fact that $\bar{\mathcal{L}}_n-\mathcal{L}_n$ and $\bar{\mathcal{G}}_r-\mathcal{G}_r$ have vanishing ${\cal O}(1)$ pieces, when we write their transformation-actions on supercoordinates and take the appropriate scalings (see \refb{sconcoord} and \refb{scale}).
Thus \refb{centch} can hold only if $c$ and $\bar c$ are large and opposite in sign (in the limit $\e\rightarrow 0$). This immediately implies that the original 2d SCFT, on which we take the nonrelativistic limit, cannot be unitary. This is of course not a problem, since there are many statistical mechanical models which are described at a fixed point by 
non-unitary CFTs.

\subsection{Nonrelativistic Superconformal Transformations in the Superspace}
In the superspace formalism, for $N=(1,1)$ supersymmetry, we introduce the fermionic coordinates $\theta, \bar{\theta}$ for the holomorphic and the antiholomorphic sectors respectively\footnote{More details can be found in \cite{qiu}.}. A superfield is a function defined on superspace, and can be expanded as a power series in $\theta, \bar{\theta}$:
\be{supfield}
\Phi(\mathcal{Z},\mathcal{\bar{Z}}) = \phi(z,\bar{z}) + \theta \psi(z,\bar{z}) + \bar{\theta}\bar{\psi}(z,\bar{z}) + \theta \bar{\theta} F(z,\bar{z}) \,,
\ee{}
where
\be{}
\mathcal{Z} \equiv (z,\theta)\,,\quad \mathcal{\bar{Z}}\equiv (\bar{z},\bar{\theta})\,.
\ee{}
The superfields correspond to irreducible representations of the Neveu-Schwarz algebra. The irreducible representations of the Ramond algebra correspond to conformal fields distinct from the superfields, which are in fact non-local (i.e., double-valued) with respect to the fermionic parts of the superfields. These are called \textit{spin fields} and they intertwine the two sectors (see, e.g., \cite{fqs1985}).

As in conformal transformations, in superconformal transformations too the unbarred and the barred parts are independent.
In superspace, the superconformal transformations corresponding to the holomorphic sector are given by:
\ben{sconcoord}
&& (z\,,\,\theta)\xrightarrow{\delta ' \mathcal{L}_n}(z - \delta '\, z^{n+1}\,,\,\theta - \delta '\, \frac{n+1}{2} z^n \,\theta)\, \crcr
&& (z\,,\,\theta)\xrightarrow{\eta\, \mathcal{G}_r} (z + \eta\,\theta\, z^{r+\frac{1}{2}}\,,\,\theta-\eta\, z^{r+\frac{1}{2}} )\,,
\een{}
where $\eta$ is an anticommuting parameter. Similarly, one can write down transformations for the antiholomorphic sector.

In terms of spacetime coordinates, $z=t+x,\,\bar{z}=t-x$. Analogously, we take linear combinations of $\theta, \bar{\theta}$ and define the new anticommuting variables:
\be{theta}
\alpha = \frac{\theta +\bar{\theta}  }{2}\,,\quad \beta = \frac{\theta -\bar{\theta} }{2}\,.
\ee

The nonrelativistic contraction corresponding to \refb{lincom} consists of taking the scalings:
\be{scale}
t \rightarrow t\,,\quad x \rightarrow \e x\,,\quad \alpha \rightarrow \alpha\,,\quad \beta \rightarrow \e \beta \,,
\ee{}
which immediately gives the coordinates in the nonrelativistic superspace transforming as:
\ben{nrcoord}
&& \delta_{\,\delta ' L_n}\lbrace t,x,\alpha,\beta \rbrace = -\delta ' \,\lbrace t^{n+1}\,,\,(n+1)\,t^n\,x\,,\frac{1}{2}(n+1) \,t^n\,\alpha\,,\,\frac{1}{2}(n+1) (\,t^n\,\beta + n\,t^{n-1}\,x\,\alpha) \rbrace\,, \, \crcr
&& \delta_{\,\delta ' M_n}\lbrace t,x,\alpha,\beta \rbrace = \delta ' \, \lbrace 0\,,\, t^{n+1}\,,\,0\,,\,\frac{1}{2}(n+1) \,t^n\,\alpha\rbrace \,, \crcr
&& \delta_{\,\eta G_r}\lbrace t,x,\alpha,\beta \rbrace = \eta \, \lbrace t^{r+\frac{1}{2}} \, \alpha\,,\,t^{r+\frac{1}{2}} \, \beta + (r+\frac{1}{2})\,t^{r-\frac{1}{2}}\,x\,\alpha\,,-t^{r+\frac{1}{2}}\,,-(r+\frac{1}{2})\,t^{r-\frac{1}{2}}\,x\rbrace \,, \crcr
&& \delta_{\,\eta H_r}\lbrace t,x,\alpha,\beta \rbrace = \eta \, \lbrace0\,,-t^{r+\frac{1}{2}}\,\alpha\,,0\,,\, t^{r+\frac{1}{2}}\rbrace \,.
\een{}

\section{Representations of the 2d SGCA}

We now turn to the representations of the 2d SGCA. In all our subsequent discussions, we consider the NS sector and hence $r,s\,\in \, \mathcal{Z}+\frac{1}{2}$ in all formulae and equations that follow. We will be guided in this by the representation theory of the SuperVirasoro algebra.

\subsection{Primary States and  Descendants}

We will construct the representations 
by considering states having definite scaling dimensions:
\begin{equation}
L_0 |\Delta \rangle = \Delta | \Delta \rangle \,.
\label{L0=Delta}
\end{equation}
Using the commutation relations \refb{sgca}, 
we obtain 
\begin{equation}
L_0 L_n | \Delta \rangle = (\Delta - n) L_n | \Delta \rangle, \quad 
L_0 M_n | \Delta \rangle = (\Delta - n) M_n | \Delta \rangle. 
\end{equation}
Then the $L_{n}, M_{n}$ with $n >0$ lower the value of the scaling
dimension, while those with $n<0$ raise it. If we demand that the
dimensions of the states be bounded from below, then we are led to
defining primary states in the theory with the properties: 
\begin{equation}
L_n|\Delta \rangle_p =0\,, \quad 
M_n|\Delta \rangle_p =0\,,\quad G_r|\Delta \rangle_p =0\,,\quad H_r|\Delta \rangle_p =0\,\mbox{\hspace{3mm}(for all $n>0$ and $r>0$)}\,.
\label{primop} 
\end{equation}
 
Since the conditions (\ref{primop})
are compatible with $M_0$ in the sense
\begin{equation}
L_n M_0 |\Delta \rangle_p = 0\,, \quad 
M_n M_0 |\Delta \rangle_p = 0\,,
\end{equation}
and also since $L_0$ and $M_0$ commute,
we may  introduce an additional label,
which we will call ``rapidity'' $\xi$:
\begin{equation}
M_0 |\Delta, \xi \rangle_p = \xi |\Delta, \xi \rangle_p\,.
\end{equation}

Starting with a primary state $|\Delta,\xi \rangle_p$ , 
one can build up a tower of operators by the action of 
$L_{-n}\,,\,M_{-n}\,,\,G_{-r}\,,\,H_{-r}$ with $n,r>0$. These will
be called the SGCA descendants of the primary. 
The primary state together with its SGCA descendants 
form a representation of SGCA. 
As in the SuperVirasoro case, we have to be 
careful about the presence of null states.
We will look at these in some detail
later in Sec.~\ref{SGCAnull}. 

The above construction is quite analogous to that of the 
relativistic 2d SCFT. 
In fact, from the viewpoint of the limit (\ref{lincom}), we see that the two labels $\Delta$ and $\xi$ are related to the conformal weights in the 2d SCFT as
\be{delxi}
\Delta=\lim_{\epsilon \to 0}
(h+\bar h)\,, \qquad  \xi= \lim_{\e\to 0} \e ({\bar{h} -h})\,,
\ee 
where $h$ and $\bar h$ are the eigenvalues of $\L_0$ and $\bL_0$, respectively. 
We will proceed to assume that such a scaling limit (as $\e \to 0$) of the 2d SCFT 
exists. In particular, we will assume that the operator-state correspondence 
in the 2d SCFT gives a similar correspondence between the states and the operators
in the SGCA\footnote{We thank the referee for emphasizing that this is an assumption we are making (without any justification). Our approach here is to go ahead with this assumption and examine whether this leads to interesting structures and whether the various algebraic considerations lead to a consistent picture.}:
\be{stateop}
\O (t,x) \leftrightarrow \O(0,0)\,|0\rangle\,, 
\ee
where $|0 \rangle$ would be the vacuum state
which is invariant under the generators 
$L_0, L_{\pm1}$, $M_0, M_{\pm1}$.  
Indeed in the rest of the paper, we will
offer several pieces of evidence that the scaling limit 
gives a consistent quantum mechanical system.

\subsection{Transformation Laws of Superprimary fields}

We consider the transformation laws of SGCA primary superfields arising from the transformation laws of primary superfields in 2d SCFT, which are given by ( following \cite{qiu} ):
\ben{sprim}
&&[\, \mathcal{L}_n , \Phi(z,\bar{z},\theta,\bar{\theta}) \,] = [\,z^{n+1} \p_z + \frac{1}{2} (n+1) z^n \theta \p_{\theta} + h(n+1) z^n\,] \Phi(z,\bar{z},\theta,\bar{\theta}) \,,\crcr
&&[\,\eta \,\mathcal{G}_r , \Phi(z,\bar{z},\theta,\bar{\theta}) \,] = \eta \,[\,z^{r+ \frac{1}{2}} (\p_{\theta} - \theta \p_z)-2h (r+\frac{1}{2})z^{r - \frac{1}{2}\,} \theta\,] \Phi(z,\bar{z},\theta,\bar{\theta}) \,;
\een
the transformations corresponding to $ \bar{\mathcal{L}}_n\,,\,  \bar {\mathcal{G}}_r  $  are given by replacing $z \to\bar z$, $\theta \to \bar{\theta}$ and $h \to \bar{h}$.
We should note here that $(h,\bar{h})$ corresponds to the conformal weights of the lowest component $\phi$ of the superfield $\Phi$ in \refb{supfield}.

Motivated from the relation (\ref{lincom}), we may 
define the transformations generated by $L_n, M_n, G_n, H_n$ as:
\ben{stras}
&&[L_n,\Phi] = \lim_{\epsilon \to 0}\, [\, \bar{\mathcal{L}}_n + \mathcal{L}_n\,,\,\Phi ]\,,
\qquad 
[M_n,\Phi]= \lim_{\epsilon \to 0}\, \e\, [\, \bar{\mathcal{L}}_n-\mathcal{L}_n\, ,\,\Phi]\,,\crcr
&& [G_r, \Phi] = \lim_{\epsilon \to 0}\, [\, \bar{\mathcal{G}}_r + \mathcal{G}_r\,,\,\Phi ]\,,
\qquad 
[H_r,\Phi]= \lim_{\epsilon \to 0}\, \e \,[\,  \bar{\mathcal{G}}_r-\mathcal{G}_r\, ,\,\Phi ]\,,
\een{}
where the superfield $\Phi$ is now a function of $\lbrace t,x,\alpha,\beta \rbrace$ and is expanded as:
\be{nsupfield}
\Phi(t,x,\alpha,\beta)= \phi_1 (t,x)+ \alpha \,\psi_1(t,x) + \beta\, \psi_2(t,x) + \alpha \beta \,\phi_2 (t,x)\,.
\ee{}

Then by taking the limits on the superspace coordinates, we obtain:
\ben{stras1}
&&[L_n\,,\Phi] = [\,t^{n+1}\,\p_t\,+\,(n+1)\,t^n\,x\,\p_x\,+\,(n+1)\,(\Delta\,t^n-n\,\xi\,t^{n-1}\,x)\,\crcr
&& \hspace{35mm} +\,\frac{1}{2}(n+1)\,\lbrace \,t^n\,(\alpha\,\p_\alpha\,+\,\beta\,\p_\beta)\,+\,n\,t^{n-1}\,x\,\alpha \,\p_\beta \rbrace \,]\,\Phi\,,\crcr
&& [M_n\,,\Phi] = [\,-t^{n+1}\,\p_x +\,(n+1)\,\xi\,t^n-\frac{1}{2} (n+1)\,t^n\,\alpha\,\p_\beta]\,\Phi\,,\crcr
&&[\eta_G\, G_r\,, \Phi] = \eta_G \,[\,t^{r+\frac{1}{2}}\,(-\alpha \p_t -\beta \p_x +\,\p_{\alpha} ) + (\,r+\frac{1}{2}\,)\,t^{r-\frac{1}{2}}\,x \, (-\alpha \p_x + \p_{\beta}) \crcr
&& \hspace{35mm} + \,2 \,(\,r+\frac{1}{2}\,)\,t^{r-\frac{1}{2}}\,(\xi \,\beta - \Delta \,\alpha) + \,2 \,(\,r^2 - \frac{1}{4}\,)\,\xi \,t^{r-\frac{3}{2}} \,x \,\alpha \,] \Phi\,,\crcr
&& [\eta_H \,H_r\,, \Phi] = \eta_H\,[\,t^{r+\frac{1}{2}}\,(\alpha \p_x - \p_\beta )-2\,(\,r+\frac{1}{2}\,)\,\xi\,t^{r-\frac{1}{2}}\,\alpha\,]\, \Phi \,.
\een
where $\eta_G, \eta_H$ are anticommuting parameters. Note that the part of the transformation laws independent of $\Delta$ and $\xi\,$, involving superspace derivatives, encodes the change due to superspace coordinate dependence of $\Phi$, and is in perfect agreement with \refb{nrcoord}.

Introducing the vacuum state $|0 \rangle$ satisfying
\ben{vacuum}
&& L_n |0 \rangle =0\,,\quad M_n |0 \rangle =0\,,\,\,\,\mbox{              (for $n\geq -1$)}\crcr
&& G_r |0 \rangle = 0\,,\quad H_r |0 \rangle = 0\,,\,\,\,\mbox{              (for $r \geq - \frac{1}{2}$)}\,,
\een{}
one immediately finds from \refb{stras1} that
\ben{desc}
G_{1 \over 2} |\phi_1\rangle = 0\,,&&\quad G_{-\frac{1}{2}}  |\phi_1\rangle = |\psi_1\rangle \,,\crcr
H_{1 \over 2} |\phi_1\rangle = 0\,,&&\quad H_{-\frac{1}{2}}  |\phi_1\rangle = -|\psi_2 \rangle\,,\crcr 
G_{-\frac{1}{2}} G_{-\frac{1}{2}}  |\phi_1\rangle =L_{-1}|\phi_1\rangle \,,
&&\quad H_{-\frac{1}{2}} H_{-\frac{1}{2}}  |\phi_1\rangle = 0\,,\crcr
H_{-\frac{1}{2}} G_{-\frac{1}{2}}  |\phi_1\rangle =M_{-1}|\phi_1\rangle - |\phi_2\rangle\,,
&&\quad G_{-\frac{1}{2}} H_{-\frac{1}{2}}  |\phi_1\rangle = M_{-1}|\phi_1\rangle + |\phi_2\rangle\,,
\een{}
where the state $|\phi_1\rangle= \phi_1(0,0)\,|0\rangle$ satisfies the conditions \refb{primop} for a primary state.

\section{Non-Relativistic Ward Identities and Correlation Functions} 

In \cite{qiu}, the two and three point functions for the 2d SCFT were found using the superspace formalism. Here we take the appropriate limits of the those correlation functions to get the SGCA correlation functions and check that these obey the Ward identities coming from the global part comprising $\lbrace L_0,\,L_{\pm1},\,M_0,\,M_{\pm 1},\,G_{\pm \frac{1}{2}},\,H_{\pm \frac{1}{2}}\rbrace$. One can solve the differential equations coming from the Ward identities to find the correlation functions directly using \refb{stras1}. However, the calculation becomes cumbersome because here one cannot use the nice property of the independence of holomorphic and antiholomorphic sectors of the SCFT. We solve the differential equations for the two point functions directly with the SGCA operators, whereas, for the three point function, we find the expression only by taking the limit of the SCFT answer.

For the sake of completeness, we state here the differential equations that an n-point function, 
\be{}
G^{(n)}_{\rm 2d\,SGCA}(\lbrace t_i,x_i,\alpha_i,\beta_i \rbrace)=\langle \Phi_1(t_1,x_1,\alpha_1,\beta_1) \,\Phi_2(t_2,x_2,\alpha_2,\beta_2) \cdots \Phi_n(t_n,x_n,\alpha_n,\beta_n) \rangle\,, \nonumber
\ee{}
should satisfy:
\ben{npt}
&& \Big[ \sum_{i=1}^n \p_{t_i} \Big ] G^{(n)}_{\rm 2d\,SGCA} =0 \,,\label{L-1}\crcr
&& \Big [ \sum_{i=1}^n \p_{x_i} \Big ] G^{(n)}_{\rm 2d\,SGCA} =0 \,,\label{M-1}\crcr
&& \Big[ \sum_{i=1}^n \lbrace t_i\p_{t_i} + x_i \p_{x_i} + \Delta_i + \frac{1}{2} (\alpha_i \p_{\alpha_i} + \beta_i \p_{\beta_i}) \rbrace \Big] G^{(n)}_{\rm 2d\,SGCA} =0 \,,\label{L0}\crcr
&& \Big[ \sum_{i=1}^n \lbrace -t_i\p_{x_i} + \xi_i - \frac{1}{2} \alpha_i \p_{\beta_i} \rbrace \Big] G^{(n)}_{\rm 2d\,SGCA} =0 \,,\label{M0}\crcr
&& \Big[ \sum_{i=1}^n \lbrace t_i^2\p_{t_i} + 2 t_i x_i \p_{x_i} + 2 (\Delta_i t_i -\xi_i x_i)+ t_i(\alpha_i \p_{\alpha_i} + \beta_i \p_{\beta_i}) + x_i \alpha_i  \p_{\beta_i} \rbrace \Big] G^{(n)}_{\rm 2d\,SGCA} =0 \,,\label{L1}\crcr
&& \Big[ \sum_{i=1}^n \lbrace -t_i^2\p_{x_i} + 2 \xi_i t_i  - t_i \alpha_i \p_{\beta_i} \rbrace \Big] G^{(n)}_{\rm 2d\,SGCA} =0 \,,\label{M1}\crcr
&& \Big[ \sum_{i=1}^n \lbrace -\alpha_i \p_{t_i} - \beta_i \p_{x_i}+ \p_{\alpha_i} \rbrace \Big] G^{(n)}_{\rm 2d\,SGCA} =0 \,,\label{G-1/2}\crcr
&& \Big[ \sum_{i=1}^n \lbrace \alpha_i \p_{x_i} - \p_{\beta_i} \rbrace \Big] G^{(n)}_{\rm 2d\,SGCA} =0 \,,\label{H-1/2}\crcr
&& \Big[ \sum_{i=1}^n \lbrace t_i( -\alpha_i \p_{t_i} -\beta_i \p_{x_i}+ \p_{\alpha_i}) + x_i (-\alpha_i \p_{x_i} + \p_{\beta_i} ) + 2 (\xi_i \beta_i - \Delta_i \alpha_i) \rbrace \Big] G^{(n)}_{\rm 2d\,SGCA} =0 \,,\label{G1/2}\crcr
&& \Big[ \sum_{i=1}^n \lbrace t_i (\alpha_i \p_{x_i} - \p_{\beta_i})-2 \xi_i \alpha_i \rbrace \Big] G^{(n)}_{\rm 2d\,SGCA} =0 \,.\label{H1/2}
\een{}
The above constraints follow from invariance under the generators $L_{-1}$, $M_{-1}$, $L_{0}$, $M_{0}$, $L_{1}$, $M_{1}$, $G_{-\frac{1}{2}}$, $H_{-\frac{1}{2}}$, $G_{1 \over 2}$ and $H_{1 \over 2}$ respectively.

\subsection{SGCA Two Point Functions}
We derive the two point functions between all components of two superfields
\be{nsupfield1}
\Phi_i(t_i,x_i,\alpha_i,\beta_i)=\phi_{i1}(t_i,x_i) + \alpha_i \, \psi_{i1}(t_i,x_i) + \beta_i \, \psi_{i2}(t_i,x_i) + \alpha_i \beta_i\, \phi_{i2}(t_i,x_i) \,,
\ee{}
with $i=1,2$. Here the lowest component fields $\phi_{i1}$ are the primary fields (see the definition \refb{primop}) and are labelled by the eigenvalues $(\Delta_i,\xi_i)$.

Here we consider the transformation rules for each component by comparing the coefficients of $\alpha^m\,\beta^n$ ( where $m,n=0,1$ ) on both sides of \refb{stras1}.

One immediately finds that the field $\phi_1(t,x)$ in \refb{stras1} has the same transformation properties under the bosonic SGCA generators as the primary fields of GCA (see eq. (4.5) and eq. (4.6) of \cite{2dgca}). Hence the $\phi_{i1}$ two point function will have the same form as derived in \cite{Bagchi:2009ca}, i.e.,
\be{2pt}
\langle \phi_{11}(t_1,x_1) \phi_{21}(t_2,x_2)\rangle = C_{12} \,\delta_{\Delta_1,\Delta_2}\, \delta_{\xi_1,xi_2} \,t_{12}^{-2\Delta_1} \,\exp(\frac {2\xi_1 x_{12}}  {t_{12}})\,,
\ee{}
where
\be{}
t_{ij}=t_i-t_j\,,\quad x_{ij}=x_i-x_j \,,
\ee{}
and $C_{12}$ is an arbitrary constant. We can take $C_{12}=1$ by choosing the normalization of the operators.

Starting from this expression, we apply the constraints coming from the fermionic generators $G_{\pm \frac {1}{2}}, H_{\pm \frac{1}{2}}$ of the global part of the SGCA to obtain all other two point functions of the superfield components, as indicated below.

Using the fact that the two point function should be a function of products of the fermionic coordinates which are Grassmann even, we immediately infer:
\be{phi-psi}
\langle \phi_{1a}\, \psi_{2b}\rangle = 0\,,\quad \langle \psi_{1a}\, \phi_{2b}\rangle = 0\,\qquad \mbox{(where $a,b=1,2$)} \,.
\ee{}
Evaluating the trivial constraint $\delta_{G_{-\frac{1}{2}}}\langle \phi_{11}\, \psi_{21}\rangle = 0$ , one gets the expression:
\be{eq1}
\langle \psi_{11}\, \psi_{21}\rangle = \p_{t_{12}} \langle \phi_{11}\, \phi_{21}\rangle = -\,\frac{2}{t_{12}}\,(\,\Delta_1 + \frac { \xi_1 x_{12}}{t_{12}}\,)\langle \phi_{11}\, \phi_{21}\rangle \,.
\ee{}
The trivial constraint $\delta_{H_{-\frac{1}{2}}} \langle \phi_{11}\, \psi_{22} \rangle = 0 $ gives:
\be{eq2}
\langle \psi_{12}\, \psi_{22}\rangle = 0\,
\ee{}
The trivial constraints $\delta_{G_{-\frac{1}{2}}} \langle \phi_{11}\, \psi_{22}\rangle = 0\,$ and $\delta_{G_{\frac{1}{2}}} \langle \psi_{12}\, \phi_{21}\rangle = 0$, on using \refb{eq2}, give the results:
\ben{}
&& \langle \phi_{12}\, \phi_{21}\rangle = 0\,,\label{eq3} \\
&&\langle \psi_{11}\, \psi_{22}\rangle =  \p_{x_{12}} \langle \phi_{11}\, \phi_{21}\rangle = \frac{2 \xi_1}{t_{12}} \langle \phi_{11}\, \phi_{21}\rangle \,.\label{eq4}
\een{}
Using $\delta_{G_{-\frac{1}{2}}} \langle \psi_{12}\, \phi_{21}\rangle = 0$ , we get:
\be{eq5}
\langle \psi_{12}\, \psi_{21}\rangle = \p_{x_{12}} \langle \phi_{11}\, \phi_{21}\rangle = \frac{2 \xi_1}{t_{12}} \langle \phi_{11}\, \phi_{21}\rangle \,.
\ee{}
Using $\delta_{G_{\frac{1}{2}}} \langle \psi_{12}\, \phi_{21}\rangle = 0$ along with \refb{eq4} and \refb{eq5}, we get:
\be{eq6}
\langle \phi_{12}\, \phi_{21}\rangle = 0\,.
\ee{}
Lastly, $\delta_{G_{\frac{1}{2}}} \langle \psi_{11}\, \phi_{22}\rangle = 0\,$, on using \refb{eq3}, \refb{eq4} and \refb{eq5}, gives:
\be{17}
\langle \phi_{12}\, \phi_{22}\rangle = \frac{4 \xi^2_1}{t_{12}^2}\langle \phi_{11}\, \phi_{21}\rangle \,.
\ee{}

Hence we find that all non-vanishing two point functions of the components of the two superfields are determined in terms of the two point function of their lowest components.

\subsection{SGCA Higher Point Functions}

Using the fact that the lowest components $\phi_{i1}$'s obey the same transformation rules as the GCA primaries under the bosonic generators of the SGCA, we conclude that all correlation functions involving these fields have the same form as one gets in the GCA case. In particular, the result derived for three point function in \cite{Bagchi:2009ca} is applicable here for $\langle \phi_{11} \phi_{21} \phi_{31}\rangle$. For the four point function of the $\phi_{i1}$'s, we can apply the same analysis as discussed in \cite{2dgca}, where one of the $\phi_{i1}$'s have a descendant null state at some level \footnote{Note that here we can have half-integer level null states. In
particular, we show in Sec.~5 that the first non-trivial null state is obtained at level $\frac{3}{2}$ and one can derive the four point function with a primary having such a descendant null state.}.  Then, as in the case of the two point function, the fermionic generators of the global part will relate the n-point function $\langle \phi_{i1}\, \phi_{i+1 1} \,\cdots \,\phi_{i+n 1} \rangle$ to the n-point functions involving arbitrary component fields of the relevant superfields $\lbrace \Phi_i,\Phi_{i+1},\cdots,\Phi_{i+n} \rbrace$.

We remind the reader that the above property follows from the fact that, in 2d CFTs and GCAs, the descendant field correlators can be derived from the primary field correlators. Here the component fields $\psi_{i1},\psi_{i2}, \phi_{i2}$ are descendants of the primary $\phi_{i1}$, as shown in \refb{desc}. The global part of the SGCA, which closes by itself and hence forms a subgroup, allows us to group these four fields into the superfield $\Phi_i$ (supermultiplet), which is nothing but an irreducible representation of the global subalgebra.

\subsection{SGCA Correlation Functions from 2d SCFT}

We now show that the above expressions for the SGCA two point functions can also be obtained by taking an appropriate scaling limit of the 2d SCFT answers. This limit requires scaling the quantum numbers of the operators as \refb{delxi}, along with the nonrelativistic limit \refb{scale} for the coordinates.

Let us first study the scaling limit of the 
two point correlator of two superfields ( see \cite{qiu} ) given by the expression
\be{2ptcft}
G^{(2)}_{\rm 2d\,SCFT} 
=\langle\, \Phi_1(\mathcal{Z}_1,\mathcal{\bar{Z}}_1)\, \Phi_2(\mathcal{Z}_2,\mathcal{\bar{Z}}_2)\,\rangle=\delta_{h_1, \,h_2} \,\delta_{{\bar h}_1, \,{\bar h}_2}  
{\tilde{z}}_{12}^{\,-2 h_1} \,\bar {\tilde{z}}_{12}^{\,-2 \bar h_1} \,,
\ee{}
where 
\ben{}
&& z_{ij} = z_i - z_j \,,\quad {\bar{z}}_{ij} = \bar z_i - \bar z_j \,,\crcr
&& {\tilde{z}}_{ij} = z_{ij} - \theta_i\,\theta_j \,,\quad \bar {\tilde{z}}_{ij} ={\bar{z}}_{ij} - \bar \theta_i\,\bar \theta_j \,.
\een{}
On scaling the above expression according to \refb{scale} and taking the limit using \refb{delxi}, it reduces to:
\ben{2ptnr}
 G^{(2)}_{\rm 2d\,SGCA}
&=& \lim_{\epsilon \to 0} \delta_{h_1,\,h_2}\,\delta_{\bar h_1,\,\bar h_2}\,\,
\lbrace t_{12}-\alpha_1 \alpha_2 + \e \,(x_{12}-\alpha_1 \beta_2 + \alpha_2 \beta_1) + \e^2 \,\beta_1 \beta_2 \rbrace ^{-2h_1} \crcr 
&& \hspace{30 mm} \times \,\lbrace t_{12}-\alpha_1 \alpha_2 + \e \,(x_{12}-\alpha_1 \beta_2 + \alpha_2 \beta_1) + \e^2 \,\beta_1 \beta_2 \rbrace ^{-2 \bar h_1} \crcr 
&& \crcr
&=& \lim_{\epsilon \to 0} \delta_{h_1,\,h_2}\,\delta_{\bar h_1,\,\bar h_2}\,\,
( t_{12}-\alpha_1 \alpha_2) ^{- 2(h_1 +\bar h_1)} \crcr 
&& \hspace{30 mm} \times \exp \lbrace
- 2 (h_1- \bar h_1)  \big(\epsilon \,\frac {( x_{12}-\alpha_1 \beta_2 + \alpha_2 \beta_1)} {(t_{12}-\alpha_1 \alpha_2)} 
+ {\cal O}(\epsilon^2)\,\big) \rbrace \,\crcr 
&& \crcr
&=& \delta_{\Delta_1,\,\Delta_2}\,\delta_{\xi_1,\,\xi_2}\,\,{\tilde{t}}_{12}^{- 2\Delta_1}
\exp \Big( \frac{2 \xi_1 {\tilde{x}}_{12}}{{\tilde{t}}_{12}} \Big)\,,
\een{}
where 
\be{}
{\tilde{t}}_{ij}=t_{ij}-\alpha_i \alpha_j \,,\quad {\tilde{x}}_{ij}=x_{ij} -\alpha_i \beta_j + \alpha_j \beta_i\,.\,\,\,
\ee{}

Now expanding the LHS $ G^{(2)}_{\rm 2d \,SGCA} \equiv \langle \Phi_1 \Phi_2 \rangle $ using \refb{nsupfield}, and comparing the coefficients of $\alpha_1^k \beta^l_1 \alpha_2^m \beta_2^n$ (for $k,l,m,n=0,1$) on both sides of \refb{2ptnr}, we get the values of all possible two point functions of the component fields. One can check that these answers exactly match with those obtained in \refb{eq3}-\refb{17}. Also, working in superfield formalism, one can check that \refb{2ptnr} satisfies the constraints coming from the global part of the SGCA using directly \refb{npt} (i.e., without considering the transformations of the component fields separately).

Another interesting point to note is the following:
Transforming the nonrelativistic superspace coordinates $\lbrace t_1,\,x_1,\,\alpha_1,\,\beta_1 \rbrace$ (using  \refb{nrcoord}) successively by $t_2 \, L_{-1} $, $-x_2\,M_{-1}$, $\alpha_2 \, G_{-\frac{1}{2}}$ and $-\beta_2\, H_{-\frac{1}{2}}\,$, we move to the point in the superspace labelled by $\lbrace {\tilde{t}}_{12}\,,\,{\tilde{x}}_{12}\,,\,\alpha_1-\alpha_2\,,\,\beta_1-\beta_2 \rbrace$. The vacuum being invariant under these global transformations, one can easily see that the two-point function should be a function of these combinations of the six coordinates $\lbrace t_i,\,x_i,\,\alpha_i,\,\beta_i \rbrace$.

A similar analysis yields the three point 
function of the SGCA from the relativistic 
three point function.
The relativistic three point function is written as:
\ben{rel3pt}
G^{(3)}_{\rm 2d\,CFT} 
& = & \langle\, \Phi_1(\mathcal{Z}_1,\mathcal{\bar{Z}}_1)\, \Phi_2(\mathcal{Z}_2,\mathcal{\bar{Z}}_2)\,\Phi_3(\mathcal{Z}_3,\mathcal{\bar{Z}}_3)\,\rangle \crcr
&=& [\,{\tilde{z}}_{12}^{\,\,h_3-h_1-h_2}\, \,\,
{\tilde{z}}_{23}^{\,\,h_1-h_2-h_3} \,\,\,
{\tilde{z}}_{31}^{\,\,h_2-h_3-h_1} 
\times\,(\textrm{antiholomorphic}) \,]\crcr
&& \,\,\times \, [\,C_{123} \, + \,\frac{\tilde{C}_{123}} { | {\tilde{z}}_{12}\,{\tilde{z}}_{23}\,{\tilde{z}}_{31} | }\,\lbrace(\theta_1\,{\tilde{z}}_{23} + \theta_2 \,{\tilde{z}}_{31}+\theta_3\,{\tilde{z}}_{12}+ \theta_1 \,\theta_2\, \theta_3)\,\times\,(\textrm{antiholomorphic})\rbrace \,]\,. \crcr
&&
\een{}
One should note that there are two arbitrary constants $C_{123}\,,\, {\tilde {C}}_{123}$ in $G^{(3)}_{\rm 2d\,CFT} $.

Taking the nonrelativistic limit, we obtain the SGCA three point function as:
\ben{nrel3pt}
G^{(3)}_{\rm 2d\,SGCA} 
& = & C_{123}\,\,
{\tilde{t}}_{12}^{\,\,\D_{3}-\D_{1}-\D_{2}}\,\, \,
{\tilde{t}}_{23}^{\,\,\D_{1}-\D_{2}-\D_{3}}\,\, \,
{\tilde{t}}_{31}^{\,\,\D_{2}-\D_{3}-\D_{1}} \nonumber \\[1mm]
&& \hspace{1.2 cm}\times \,
\exp \lbrace
\frac {(\xi_1+ \xi_2 -\xi_3)\,{\tilde{x}}_{12} }  {{\tilde{t}}_{12}}
+ 
\frac{ (\xi_2+ \xi_3 -\xi_1)\, {\tilde{x}}_{23} }  {{\tilde{t}}_{23}}  
+ 
\frac{ (\xi_1+ \xi_3 -\xi_2)\,{\tilde{x}}_{31} }  {{\tilde{t}}_{31}} \rbrace \,.\crcr
&&
\een{}
Again, one can check that \refb{nrel3pt} satisfies the differential equations \refb{npt}. Comparing the coefficients of the parts involving no fermionic coordinates $\lbrace \alpha_i,\beta_i \rbrace$ on both sides, we find that $\langle \phi_{11} \phi_{21} \phi_{31}\rangle$ is exactly what was derived in \cite{Bagchi:2009ca}, and this is what one should get following the discussion in Sec.~4.2.

Here we note that the contribution from the part multiplying $\tilde{C}_{123}$ in \refb{rel3pt} is zero in the nonrelativistic limit. However, it may so happen that $\tilde{C}_{123}$ scales in a manner so as to give a finite contribution in combination with the $\mathcal{O}(\e)$ terms. This cannot be ascertained just from the relativistic answer. We need to examine whether the second part survives in the nonrelativistic limit by verifying whether it is possible to satisfy \refb{npt} by keeping the $\mathcal{O}(\e)$ terms. Examining the three point functions of the various component fields with the extra terms, we find that \refb{npt} is not satisfied. Hence we conclude that $G^{(3)}_{\rm 2d\,SGCA}$ is completely specified by \refb{nrel3pt}.

\section{SGCA Null Vectors}
\label{SGCAnull}
Just as in the representation of the SuperVirasoro algebra, 
we will find that there are null states in the SGCA tower built on a
primary  $|\Delta, \xi\rangle$ for special values of $(\Delta,\xi)$.  
These are states which are orthogonal to all states in the tower
including itself.

We can find the null states at a given level by writing the most general
state at that level as a combination of the  $L_{-m}, M_{-n}, G_{-\frac{r}{2}},H_{-\frac{s}{2}}$'s and their products (for $m,n,r,s>0$) acting on the SGCA primary, and then imposing the condition that all the positive modes $L_m, M_n,G_{\frac{r}{2}},H_{\frac{s}{2}}$ (with $m,n,r,s>0$) annihilate this state. This will give conditions that fix the relative coefficients in the
linear combination as well as give a relation between  
$\Delta,\xi$ and the central charges $C_1\,, C_2$. This procedure will give us null states which are primaries and descendants at the same time. These are called ``singular vectors''.

In this context, we would like to mention that for $C_2=0\,$, since the vacuum state satisfies \refb{vacuum}, all states of the form $M_{-n}|0\rangle$ and $H_{-s}|0\rangle$ (for $n,s > 0$) are null states, as their correlation functions with other primaries and secondaries will vanish. Similarly, for $C_1=C_2=0\,$, $\,L_{-m}|0\rangle$ and $G_{-r}|0\rangle$ (for $m,r >0 $) will also be null states.\footnote{Note that these null states are not highest-weight states, and hence not singular vectors. We thank the referee for emphasizing this point.} Hence, for these special cases, the correlation functions will satisfy much stronger constraints as stated below\footnote{We would like to thank Ashoke Sen for pointing this out.}:

(a) For $C_2=0$, the correlators are invariant under the generators $M_{-n}$ and $H_{-s}\,$, resulting in the equations:
\ben{C2eq0}
&& \Big[ \sum_{i=1}^k \Big \lbrace -\frac{1}{t_i^{n-1}}\,\p_{x_i} -\frac{(n-1) \,\xi_i} {t_i^n}  + \frac{(n-1)\, \alpha_i}{2\, t_i^n}\, \p_{\beta_i} \Big \rbrace \Big] \,G^{(k)}_{\rm 2d\,SGCA} =0 \,,\label{M-n}\\
&& \Big[ \sum_{i=1}^k \Big \lbrace \frac{1} {t_i^{s-\frac{1}{2}}}\, (\alpha_i \p_{x_i} - \p_{\beta_i})+ 2 \,(s-\frac{1}{2})\,\frac {\xi_i\, \alpha_i}{t_i^{s+\frac{1}{2}}} \Big \rbrace \Big] \,G^{(k)}_{\rm 2d\,SGCA} =0 \,.\label{H-s}
\een{}
Acting on the two point function $G^{(2)}_{\rm 2d\,SGCA}$, these constraints give the condition $\xi=0\,$, which removes the spatial and $\beta$ dependence of the correlators\footnote{This follows from the fact that all $x$ and $\beta$ dependence arises in combination with the $\xi$ dependence so as to survive the nonrelativistic limit.}.

(b) For $C_1=C_2=0$, the correlators are invariant under the generators $L_{-m}\,$, $G_{-r}\,$, $M_{-n}$ and $H_{-s}\,$, resulting in the equations:
\ben{C2C1eq0}
&& \Big[ \sum_{i=1}^k \Big\lbrace \frac{1}{t_i^{m-1}}\,\p_{t_i}-\frac{(m-1)\,x_i}{t_i^m}\,\p_{x_i}-\frac{m-1}{t_i^m}\,\Big(\Delta_i +\frac{m\,\xi_i\,x_i}{t_i}\Big)\,\crcr
&& \hspace{15mm} -\frac{m-1}{2\,t_i^m}\,\Big(\alpha_i\,\p_{\alpha_i}\,+\,\beta_i\,\p_{\beta_i}-\frac{m\,x_i\,\alpha_i}{t_i} \,\p_{\beta_i} \Big) \Big \rbrace \Big] \,G^{(k)}_{\rm 2d\,SGCA} =0 \,,\label{L-m}\\
&& \Big[ \sum_{i=1}^k \Big\lbrace \frac{1}{t_i^{r-\frac{1}{2}}}\,(-\alpha_i \p_{t_i} -\beta_i \p_{x_i} +\,\p_{\alpha_i} ) -(r-\frac{1}{2})\,\frac {x_i} {t_i^{r+\frac{1}{2}}}\, (-\alpha_i \p_{x_i} + \p_{\beta_i}) \crcr
&& \hspace{15mm} -2 \,(r-\frac{1}{2})\,\frac{1}{t_i^{r+\frac{1}{2}}}\,(\xi_i \,\beta_i - \Delta_i \,\alpha_i) +\,2 \,(r^2 - \frac{1}{4})\,\frac{\xi_i  \,x_i \,\alpha_i}{t_i^{r+\frac{3}{2}}} \Big \rbrace   \Big]\,G^{(k)}_{\rm 2d\,SGCA} =0\,,\label{G-r}\nonumber \\
\een{}
in addition to \refb{M-n} and \refb{H-s}. Acting on the two point function $G^{(2)}_{\rm 2d\,SGCA}$, these constraints give the condition $\xi=\Delta=0\,$, which simply means that there is no primary in the theory except the vacuum state.

Hence, these sectors are quite trivial, and in all discussions that follow, we will assume that at least $C_2 \neq 0$ .

\subsection{The Intrinsic SGCA Analysis}
At level $\frac{1}{2}$, we can consider a general state $(a\,G_{- \frac{1} {2}}+b\, H_{-\frac {1} {2}})\,|\Delta, \xi\rangle$. One can check that we get two linearly independent null states: $ G_{-\frac {1} {2}}|\Delta=0, \xi=0\rangle$ and $ H_{-\frac {1}{2}}|\Delta, \xi=0\rangle\,$.

At level one, we have the general state $(a\, L_{-1} +\, b\, M_{-1} + \,c\, G_{- \frac{1} {2}}H_{- \frac{1} {2}})\,|\Delta, \xi\rangle$ (note that this is the most general linear combination of the lowering operators at this level, remembering the relation $\lbrace G_{- \frac{1} {2}} , H_{- \frac{1} {2}} \rbrace = 2 M_{-1}\, $). 
It is easy to check that one has three linearly independent null states given by $ L_{-1}|\Delta=0, \xi=0\rangle$ , $ M_{-1}|\Delta, \xi=0\rangle$ and $ G_{-\frac{1}{2}} H_{-\frac{1}{2}}|\Delta, \xi=0\rangle$.

At level $\frac {3}{2}$, things are a little more non-trivial. Let us consider the most general level $\frac {3}{2}$ state of the form 
\be{lev3by2}
|\chi \rangle = (a \,G_{-\frac{3}{2}}+\,b\,L_{-1} G_{-\frac{1}{2}}+\,c\,M_{-1}G_{-\frac{1}{2}}+\,d\,H_{-\frac{3}{2}}+\,e\,L_{-1} H_{-\frac{1}{2}}+\,f\,M_{-1}H_{-\frac{1}{2}})\,|\Delta, \xi\rangle \,.
\ee{}
We now impose the conditions that $G_{\frac{1}{2},\frac{3}{2}},H_{\frac{1}{2},\frac{3}{2}}, L_{1}, M_{1}$ annihilate this state\footnote{This is sufficient as the annihilation condition for all the other higher level positive modes are then automatically satisfied.}, using \refb{sgca}. This gives us the following set of conditions: 
\ben{null-state-eqn}
&& \xi \,[\,2a+(1+2\Delta)b + 2 \xi e \,] = 0\,,\crcr
&& \Delta\,[\,2a+(1+2\Delta)b + 2 \xi e \,]+\xi\,[\,(1+2\Delta)c + 2d +e+2 \xi f\,]=0\,,\crcr
&& \xi^2 \,b = 0\,,\label{3}\crcr
&& (2 \Delta+1)\xi b+ 2 \xi(a + \xi c)=0\,,\crcr
&& (\Delta + 4 C_1)a+ 2 \Delta b+(\xi + 4 C_2)d+ 2(c+e) \xi=0\,\label{5},\crcr
&& (\xi + 4 C_2)a + 2 \xi b=0\,\label{6},\crcr
&& \xi \,[\,2a + (1+2\Delta) b +  2 \xi c\,] = 0\,,\crcr
&& \Delta \, [\,2a + (1+2\Delta) b +  2 \xi c \,] + \xi\, [\,c+2d + (1+ 2 \Delta)e + 2 \xi f\,] = 0\,
\een{}
We will now separately consider the two cases where $\xi \neq 0$ and $\xi=0$.

For the case $\xi \neq 0$, we get the conditions: $b=0\,$, $c=e=-\frac{a}{\xi}\,$, and $f=\frac{ (\Delta + 1)a}{\xi^2}-\frac{d}{\xi} \,$. Now we have two further options: either $a=0$ or $a \neq 0\,$.

For $a=0\,$, to get a non-trivial solution, we must have $\xi=-4C_2\,$, and the null state is of the form:
\be{null1}
|\chi^{(1)} \rangle =(H_{-\frac{3}{2}}-{1 \over \xi}M_{-1} H_{-\frac{1}{2}})\,|\Delta, \xi\rangle \,.
\ee 

For $a\neq 0\,$, we are led to the following consistency conditions: $\xi=-4C_2$ and $\Delta = 4(1-C_1)\,$. In this case, both $a$ and $d$ can be arbitrary and all other coefficients are determined in terms of these. However, by taking a suitable linear combination with $|\chi^{(1)} \rangle\,$, we can choose $d=0$, and then we get another null state of the form:
\be{null2}
|\chi^{(2)} \rangle =[\,G_{-\frac{3}{2}}-{1 \over \xi}M_{-1} G_{-\frac{1}{2}} -{1 \over \xi}L_{-1} H_{-\frac{1}{2}}+\frac{ (\Delta + 1)}{\xi^2}M_{-1} H_{-\frac{1}{2}}\,]\,|\Delta, \xi\rangle \,.
\ee{}

For the case $\xi = 0$, since $C_2 \neq 0$, we must have $a=0\,$, $\Delta (2 \Delta +1) b =0$, $d=- \frac{\Delta \, b}{2 C_2}\,$, and $c\,,\,e\,,\,f$ are undetermined. 
For $\Delta \neq -\frac{1}{2}\,,0\,$ in general, we therefore get three null states: $G_{-\frac{1}{2}} M_{-1}\, |\Delta, \xi=0 \rangle \,$, $L_{-1} H_{-\frac{1}{2}} \, |\Delta, \xi=0 \rangle $ and $M_{-1} H_{-\frac{1}{2}} \, |\Delta, \xi=0 \rangle \,$.
For $\Delta=0\,$, we also obtain $d=0\,$, and in this case $b$ is also undetermined. Hence, by taking suitable linear combinations with the three null states for $b=0\,$, we get a new null state of the form $L_{-1} H_{-\frac{1}{2}}\,  |\Delta=0, \xi=0 \rangle $.
For $\Delta= -\frac{1}{2}\,$, we also have $d=\frac{b}{4 C_2}\,$, and again $b$ is also undetermined. Taking appropriate linear combinations with the three states for $b=0\,$, we get a new null state of the form:
\be{null3}
|\chi^{(3)} \rangle =(L_{-1} G_{-\frac{1}{2}} + \frac{1}{4 C_2} H_{-\frac{3}{2}})\,|\Delta=-\frac{1}{2}, \xi=0\rangle \,.
\ee 

Crucially, we note that all the above null states for $\xi=0$, except $|\chi^{(3)} \rangle\,$, are descendants of the level $\frac{1}{2}$ and level $1$ null states.

\subsection{SGCA Null Vectors from 2d SCFT}

If we want to examine the SGCA null states at a general level, we would have to 
perform an analysis similar to that in the SuperVirasoro representation theory. A cornerstone of this analysis is the Kac determinant which gives the values of the weights of the SuperVirasoro Primaries $h\, (\bar h)$ for which the matrix of inner products at a given level has a zero eigenvalue. For the NS algebra, this determinant is given by (found by Kac \cite{Kac1978}):
\be{kd}
\mbox{det} M_{(l)} = const. \prod (h - h_{p,q}(c))^{P_{NS}(l-\frac{p\,q}{2})}\,,
\ee
where the product runs over positive integers $p,q$ with $ \frac{p\,q}{2}\leq l$ and $|p-q|$ even. Here $P_{NS}(k)$ is the number of states, arising from a ground state, at level $k$:
\be{}
\sum_{k=0}^{\infty}\,\dfrac{1\,+\,t^{k-\frac{1}{2}}}{1\,-\,t^k}\,.\nonumber
\ee{}

The functions $h_{p,q}(c)$ can be expressed in a variety of ways. One convenient representation is:
\ben{h-rep1}
h_{p,q}(c) &=& h_0 + {1 \over 4}(p\, \alpha_+ + q\, \alpha_-)^2\,, \\
h_0 &=& {1 \over {16}}(c-1)\,, \\
\alpha_{\pm} &=& {\sqrt{1-c} \pm \sqrt{9 -c} \over {4}}\,.
\een
One can write a similar expression for the antiholomorphic sector. 
The values $h_{p,q}$ are the ones for which we have zeroes of the determinant and hence null vectors (and their descendants).

One could presumably generalise our analysis for SGCA null vectors at level $\frac{3}{2}$ and directly obtain the SGCA determinant at a general level. This would give us a relation for $\Delta$ and $\xi$ in terms of $C_1, C_2$ for which there are null states, generalising the results obtained at level $\frac{3}{2}\,$. However, instead of a direct analysis, here we will simply take the non-relativistic limit of the Kac formula and see whether one obtains sensible expressions for the  $\Delta$ and $\xi$ on the SGCA side. 

In taking the non-relativistic limit, $C_2$ is chosen to be positive. Therefore (from \eq{centch}) we need to take 
$c \ll -1$ and $\bar c \gg 1$ as $\e \to 0\,$. We then find
\ben{}
h_{p,q} &=& \frac{C_2}{4 \e} \,(p^2-1)  + {1 \over 16} [-1+5p^2-4pq-4 C_1 (p^2-1)\,] + {\cal O}(\epsilon)\,,\label{hh1} \\
\bar{ h}_{p',q'} &=& -\frac{C_2}{4 \e}\,({p'}^2-1)  + {1 \over 16} [-1+5p'^2-4p'q'-4 C_1 (p'^2-1)\,] 
+ {\cal O}(\epsilon)\,.
\label{hh2}
\een{}
Using \refb{delxi} and taking $p=p'$ \footnote{Requiring that $\Delta$ should not have a 
$\frac{1}{\epsilon}$ piece immediately implies that $p=p'$.\label{r=r's=s'}} 
\begin{eqnarray}
\Delta_{p(q,q')} &=& 
\lim_{\epsilon \to 0}\, (h_{p,q} + \bar h_{p,q'}) 
= -{1 \over 2}C_1 \,(p^2-1) + {1 \over 8}\, [\,5 p^2 -2p(q+q') -1\,] \,,
\label{delta}\\
\xi_{p(q,q')} &=& - \lim_{\e\to 0}\, {\e}\,(h_{p,q} - \bar h_{p,q'}) 
= -{1 \over 2} C_2\,(p^2-1)\label{xi}\,.
\end{eqnarray}

However, we would like to caution the reader that this nonrelativistic limit of the Kac formula does not give us all the null states of the SGCA (see the following subsection).

In the following discussion, we will focus on the null vectors at level $\frac{3}{2}$. The null vector at level $\frac{3}{2}$ in a SuperVirasoro tower is given by (see \cite{qiu})
\be{relnull}
|\chi_L \rangle = (\,{\mathcal{G}}_{-\frac{3}{2}} + \eta\, {\mathcal{L}}_{-1}{\mathcal{G}}_{-\frac{1}{2}}\,)\, 
|h \rangle \otimes |\bar h \rangle \,,
\ee
with
\be{nullcond}
\eta =  - {2 \over 2h+1}\,,
\ee
\be{nullcond2}
h = {1 \over 4}
\Big \lbrace 3 - c \pm \sqrt{(1-c)(9-c)} \Big\rbrace \,,
\ee
where the positive and negative signs before the square root correspond to the primaries of conformal weights $h_{3,1}$ and $h_{1,3}\,$, respectively (see \refb{h-rep1}).
One has a similar null state for the antiholomorphic SuperVirasoro
obtained by replacing
${\mathcal{L}}_n \to \bar {\cal L}_n\,$, ${\mathcal{G}}_r \to {\bar{\mathcal{G}}}_r\,$, $h \to \bar{h}$
and $c \to \bar c\,$.

For $h=h_{3,1}$ and $\bar{h}={\bar{h}}_{3,1}$, we get 
\be{nreldelxi}
\xi = -4 C_2\,, \qquad \Delta = 4(1-C_1)\,.
\ee
These are precisely the relations we obtained in the previous section if we require the existence of both the SGCA null states $|\chi^{(1)} \rangle,\, |\chi^{(2)} \rangle$ at level $\frac{3}{2}$.

These states themselves can be obtained by taking the nonrelativistic limit on appropriate combinations of the relativistic null vectors $|\chi_L \rangle$ and its antiholomorphic counterpart $|\chi_R\rangle$.
Consider
\be{nullplus}
|\chi^{(1)} \rangle = \lim_{\e\to 0} {\epsilon}\,(- |\chi_L \rangle  + |\chi_R \rangle\,) \,, \qquad |\chi^{(2)} \rangle = \lim_{\e \to 0}\,  (\,|\chi_L \rangle  + |\chi_R \rangle\,) .
\ee
From the expressions \eq{delxi}, we obtain $\eta={2\epsilon\over \xi}(1+{(\Delta+1)\,\epsilon\over \xi})$
and $\bar{\eta}= -{2\epsilon\over \xi}(1-{(\Delta+1)\,\epsilon\over \xi})$ upto terms of order $\epsilon^2\,$. 
Substituting this into \eq{nullplus} and using the relations \eq{lincom}, we obtain
\ben{nulllim}
&& |\chi^{(1)} \rangle =(H_{-\frac{3}{2}}-{1 \over \xi}M_{-1} H_{-\frac{1}{2}})\,|\Delta, \xi\rangle \,, \cr
&& |\chi^{(2)} \rangle =\Big \lbrace\,G_{-\frac{3}{2}}-{1 \over \xi}M_{-1} G_{-\frac{1}{2}} -{1 \over \xi}L_{-1} H_{-\frac{1}{2}}+\frac{ (\Delta + 1)}{\xi^2}M_{-1} H_{-\frac{1}{2}}\,\Big \rbrace\,|\Delta, \xi\rangle \,,
\een
which are exactly what we found from the intrinsic SGCA analysis  in \eq{null1} and \eq{null2}.

For the case $h=h_{1,3}$ and $\bar{h}=\bar{h}_{1,3}\,$, we find that $\Delta_{1(3,3)}=-1$ and $\xi_{1(3,3)}=0\,$. This is also easily seen to correspond to the null states constructed in Sec.~5.1 for $\xi=0$ and $\Delta \neq -\frac{1}{2}\,,0\,$ (which we have seen are descendants of level one null states).

A point to observe here is that the expansion of relativistic null state expressions (such as \refb{relnull}) in powers of $\epsilon$ gives us nonrelativistic null states when we consider only the coefficients of the first two lowest powers of $\epsilon$.\footnote{In fact this is true for any expression/result of the relativistic theory, from which we want to extract the corresponding nonrelativistic analogue. This directly follows from the fact that we have obtained the nonrelativistic algebra by retaining only the ${\cal O}({1\over \e})$ and $\mathcal{O} (1)$ terms of the relativistic algebra.} Also, one should consistently expand $h$ and $\bar{h}$ only upto $\mathcal{O} (1)$ (and not beyond) while considering any such expression, because of the definition of the nonrelativistic generators in \refb{lincom}.

\subsection{Discussion on SGCA Null States Not Obtained from SCFT Null States}

We would like to point out that though we find the limiting process gives answers consistent with the intrinsic SGCA analysis, working purely within SGCA, we get some null states which are not obtained in the SCFT case. These extra null states are not initally null in SCFT, but become null in the nonrelativistic scaling limit. We list such null states obtained at level $3 \over 2$:

\textbf{($i$)} $|\chi^{(1)} \rangle$ in \refb{null1} has $\xi=-4C_2$ but no restriction on $\Delta$. On the other hand, $|\chi^{(1)} \rangle$ obtained in \refb{nulllim} from SCFT null states, has $\Delta = 4(1-C_1)$ in addition to $\xi=-4C_2$. This clearly shows that we have more null states for $\xi=-4C_2$ from the intrinsic SGCA analysis.

\textbf{($ii$)} $|\chi^{(3)} \rangle$ in \refb{null3} descends from a state 
\be{}
\Big \lbrace {\bar{\mathcal{G}}}_{-3/2}\, -\,\mathcal{G}_{-\frac{3}{2}}\,+\, \frac{1}{2}\, (\bar{c} -c )\, (\,{\bar{\mathcal{L}}}_{-1} {\bar{\mathcal{G}}}_{-\frac{1}{2}}\, +\, \mathcal{L}_{-1}  \mathcal{G}_{-\frac{1}{2}}\,+\, {\bar{\mathcal{L}}}_{-1}  \mathcal{G}_{-\frac{1}{2}}\, +\, \mathcal{L}_{-1} {\bar{\mathcal{G}}}_{-1/2}\, ) \Big \rbrace |h= -\frac{1}{4}, \bar{h} = -\frac{1}{4} \rangle \nonumber
\ee{} 
on the SCFT side, which is not null. This is because, while analysing null state conditions, we never take linear combinations of descendants having different ${\mathcal{L}}_0$ and ${\bar{\mathcal{L}}}_0$ eigenvalues. On the other hand, descendant states in SGCA are eigenstates of $L_0\,$, but not necessarily of $M_0\,$. Hence we get a valid null state from the above state in SCFT, after the limiting process.

Hence we conclude that within the SGCA framework, we get more constraints arising from the differential equations involving the extra null states, over and above those resulting from SCFT. This means we get new fusion rules involving the primaries corresponding to these null states.\footnote{We discuss these issues a bit more elaborately in the concluding remarks, where we also mention the future directions we would like to follow to get a better understanding.}

\section{Differential Equations for SGCA Correlators from Null States}
\label{DiffEqGCA}
The presence of the null states gives additional relations between correlation functions which is at the heart of the solvability of relativistic (rational) (super)conformal field theories. To obtain these relations one starts with the differential operator realisations ${\hat{ \mathcal{L}}}_{-n}$ and ${\hat{ \mathcal{G}}}_{-r}$ of $\mathcal{L}_{-n}$ and $\mathcal{G}_{-r}$ respectively (with $n,r > 0$). 
Thus one has
\ben{hatlg}
\langle \Phi_k(\mathcal{Z}_k,\bar{ \mathcal{Z}}_k)\, \cdots\, \Phi_2(\mathcal{Z}_{2},\bar{ \mathcal{Z}}_{2}) \lbrace\mathcal{L}_{-n}\,\Phi_1(0,0)\rbrace \rangle
&=& {\hat {\mathcal{L}}}_{-n} \langle  \Phi_k(\mathcal{Z}_k,\bar{ \mathcal{Z}}_k)\, \cdots\, \Phi_2(\mathcal{Z}_{2},\bar{ \mathcal{Z}}_{2})\Phi_1(0,0)\rangle,\crcr
\langle \Phi_k(\mathcal{Z}_k,\bar{ \mathcal{Z}}_k)\, \cdots\, \Phi_2(\mathcal{Z}_{2},\bar{ \mathcal{Z}}_{2}) \lbrace\mathcal{G}_{-r}\,\Phi_1(0,0)\rbrace \rangle
&=& {\hat {\mathcal{G}}}_{-r} \langle  \Phi_k(\mathcal{Z}_k,\bar{ \mathcal{Z}}_k)\, \cdots\, \Phi_2(\mathcal{Z}_{2},\bar{ \mathcal{Z}}_{2})\Phi_1(0,0) \rangle ,\nonumber
\een{}
where
\ben{}
\hat {\mathcal{L}}_{-n} &=& \sum_{i=2}^k
\left\{
{(n-1) h_i\over z_i^n} + \frac{n-1}{2}\,\frac{\theta_i}{z_i^n} \,\p_{\theta_i}
-{1 \over z_i ^{n-1}}\, \partial_{z_i}
\right\}\,,\label{calL(-n)}\\
\hat {\mathcal{G}}_{-r} &=& \sum_{i=2}^k
\mbox{sign}_i\,\left\{
{(2r-1) h_i\over z_i ^{r+\frac{1}{2}}} + \frac{1}{z_i^{r-\frac{1}{2}}}\,(\, \p_{\theta_i}-\theta_{i}\,\p_{z_i}\, )
\right\}\,,\label{calG(-r)}
\een{}
where sign$_i$ is $+1$ and $-1$ for bosonic and fermionic superfields respectively.
One can write analogous expressions for the antiholomorphic sector.

For the SGCA also we can construct such operators. Firstly, we derive the expressions entirely from the SGCA side. 

Let us assume that we have a null state at a level $l$, which is a descendant of (the lowest component of) the primary superfield $\Phi_1(t_1,x_1,\alpha_1,\beta_1)$, represented as $f(\lbrace L_{-n}, M_{-m},G_{-r},H_{-s}\rbrace)\, \Phi_{1}(0,0,0,0)\,|0\rangle$\footnote{Note that $\Phi_{1}(0,0,0,0)\,|0\rangle=\phi_{11}(0,0)\,|0\rangle$ .}, where $f$ is the appropriate linear combination of the products of the SGCA generators (with $n,m,r,s > 0$ and the level adding up to $l$) such that the null state conditions are satisfied. Since the null states are orthogonal to all states, we have the condition:
\be{null-ortho}
\langle 0|\,\Phi_k(t_k,x_k,\alpha_k,\beta_k)\,\cdots\,\Phi_2(t_2,x_2,\alpha_2,\beta_2)\,\big[\,f(\lbrace L_{-n}, M_{-m},G_{-r},H_{-s}\rbrace) \,\Phi_1(0,0,0,0)\,\big]\,|0\rangle =0\,.\nonumber
\ee{}
Using \refb{stras1} and the fact that $L_{-n}\,,\, M_{-m}\,,\,G_{-r}\,,\,H_{-s}$ annihilate $\langle 0 |$, we commute $f$ past all the $\Phi_i$'s and obtain the expression:
\be{}
f(\lbrace \hat{L}_{-n}, \hat{M}_{-m},\hat{G}_{-r},\hat{H}_{-s}\rbrace) \,\langle 0|\,\Phi_k(t_k,x_k,\alpha_k,\beta_k)\,\cdots\,\Phi_2(t_2,x_2,\alpha_2,\beta_2)\,\Phi_1(0,0,0,0)\,|0\rangle =0\,,\nonumber
\ee{}
where the differential operators acting on the correlation function are given by:
\ben{null-diff}
\hat{L}_{-n} &=& -\sum_{i=2}^k\Big\lbrace \frac{1}{t_i^{n-1}}\,\p_{t_i}-\frac{(n-1)\,x_i}{t_i^n}\,\p_{x_i}-\frac{n-1}{t_i^n}\,\Big(\Delta_i +\frac{n\,\xi_i\,x_i}{t_i}\Big)\,\crcr
&& \hspace{15mm} -\frac{n-1}{2\,t_i^n}\,\Big(\alpha_i\,\p_{\alpha_i}\,+\,\beta_i\,\p_{\beta_i}-\frac{n\,x_i\,\alpha_i}{t_i} \,\p_{\beta_i} \Big) \Big \rbrace \,,\crcr
\hat{M}_{-m} &=&  - \sum_{i=2}^k \Big \lbrace -\frac{1}{t_i^{m-1}}\,\p_{x_i} -\frac{(m-1) \,\xi_i} {t_i^m}  + \frac{(m-1)\, \alpha_i}{2\, t_i^m}\, \p_{\beta_i} \Big \rbrace \,,\crcr
\hat{G}_{-r} &=& - \sum_{i=2}^k \mbox{sign}_i\,\Big\lbrace \frac{1}{t_i^{r-\frac{1}{2}}}\,(-\alpha_i \p_{t_i} -\beta_i \p_{x_i} +\,\p_{\alpha_i} ) -(r-\frac{1}{2})\,\frac {x_i} {t_i^{r+\frac{1}{2}}}\, (-\alpha_i \p_{x_i} + \p_{\beta_i}) \crcr
&& \hspace{15mm} -2 \,(r-\frac{1}{2})\,\frac{1}{t_i^{r+\frac{1}{2}}}\,(\xi_i \,\beta_i - \Delta_i \,\alpha_i) +\,2 \,(r^2 - \frac{1}{4})\,\frac{\xi_i  \,x_i \,\alpha_i}{t_i^{r+\frac{3}{2}}} \Big \rbrace  \,,\crcr
\hat{H}_{-s} &=& - \sum_{i=2}^k \mbox{sign}_i\,\Big \lbrace \frac{1} {t_i^{s-\frac{1}{2}}}\, (\alpha_i \p_{x_i} - \p_{\beta_i})+ 2 \,(s-\frac{1}{2})\,\frac {\xi_i\, \alpha_i}{t_i^{s+\frac{1}{2}}} \Big \rbrace    \,,
\een{} 
where once again we note that the factor sign$_i$ is necessary to account for the minus sign when commuting $f$ through a fermionic superfield\footnote{However, the reader should note that, though not stated explicitly, we have assumed correlation functions of bosonic superfields everywhere in this paper.}.

It follows directly from \refb{lincom} that expanding the operators
$\hat{ \mathcal{L}}_{-n}$ and $\hat {\mathcal{G}}_{-r}$ as 
\ben{}
\hat {\mathcal{L}}_{-n} 
&=& \epsilon^{-1} \hat{ \mathcal{L}}_{-n}^{(-1)} +\hat {\mathcal{L}}_{-n}^{(0)} +{\cal O}(\epsilon)\,,\crcr
\hat {\mathcal{G}}_{-r} 
&=& \epsilon^{-1} \hat{ \mathcal{G}}_{-r}^{(-1)} +\hat {\mathcal{G}}_{-r}^{(0)} +{\cal O}(\epsilon)\,,\nonumber
\een{}
(and similarly for the antiholomorphic part), we get expressions for the differential operators $\hat M_{-n}$ , $\hat L_{-n}$ , $\hat{H}_{-r}$ and $\hat{G}_{-r}$ which match exactly with \refb{null-diff}.

Therefore, correlation functions involving an SGCA descendant of a primary field are given in terms of the correlators of the primaries by the action of the corresponding differential operators $\hat M_{-n}$ , $\hat L_{-n}$ , $\hat{H}_{-r}$ and $\hat{G}_{-r}$. 

Now we will study the consequences of having null states at level $\frac{3}{2}$. We will consider the two null states  $|\chi^{(1)} \rangle \,, \,|\chi^{(2)} \rangle$ of Sec.~5.1, or rather correlators involving the corresponding fields $\chi^{(1,2)}(t_1,x_1)$. Setting the null state and thus its correlators to zero gives rise to differential equations for the correlators involving the
primary superfield $\Phi_{\Delta_1,\xi_1}(t_1,x_1,\alpha_1,\beta_1)$\footnote{Note that $\chi^{(1,2)}(t_1,x_1)$ is the descendant of the lowest component of $\Phi_{\Delta_1,\xi_1}$.} with other fields. Using the forms
 \eq{null1} and \eq{null2}, we find that the differential equations take the
 form 
\ben{}
&&(\hat{H}_{-\frac{3}{2}}-{1 \over \xi} \hat{M}_{-1} \hat{H}_{-\frac{1}{2}})
\,\langle \,\Phi_k(t_k,x_k,\alpha_k,\beta_k)\,\cdots\,\Phi_2(t_2,x_2,\alpha_2,\beta_2)\,\Phi_1(0,0,0,0)\, \rangle =0\,, 
\label{nulleq1} \crcr
&& \\
&& \Big[\,\hat{G}_{-\frac{3}{2}}-{1 \over \xi_1} \hat{M}_{-1} \hat{G}_{-\frac{1}{2}} -{1 \over \xi_1} \hat{L}_{-1} \hat{H}_{-\frac{1}{2}} \crcr
&& \hspace{3 mm}+\,\,\frac{ (\Delta_1 + 1)}{\xi_1^2}\, \hat{M}_{-1} \hat{H}_{-\frac{1}{2}}\,\Big]\,\langle 
\,\Phi_k(t_k,x_k,\alpha_k,\beta_k)\,\cdots\,\Phi_2(t_2,x_2,\alpha_2,\beta_2)\,\Phi_1(0,0,0,0)\, \rangle =0\,, 
\label{nulleq2}\crcr
&&
\een{}
with $\hat M_{-n}$ , $\hat L_{-n}$ , $\hat{H}_{-r}$ and $\hat{G}_{-r}$ as given in \refb{null-diff}.

\section{SGCA Fusion Rules}

Analogous to the relativistic case (see \cite{berhadsky} and \cite{sotkov}), 
we can derive ``\textit{Fusion rules}",
\be{fus}
[\Phi_1] \times [\Phi_2] \simeq \sum_{f} [\Phi_f] \,\, ,\nonumber 
\ee{}
for the SGCA superconformal families, that determine which families $[\Phi_f]$ have their primaries and descendants occurring in an OPE of any two members of the families $[\Phi_1]$ and $[\Phi_2]$. Here we have denoted a family $[\Phi_i]$ by the corresponding primary superfield $\Phi_i$.

We illustrate how the fusion rules can be obtained for the families $[\Phi_{\Delta_1 , \xi_1}]$ and $[\Phi_{\Delta_2 , \xi_2}]$, where both fields are members of the the nonrelativistic limit of the Kac table as specified by \refb{hh1} and \refb{hh2}. 
As mentioned in footnote \ref{r=r's=s'}, we need to take $p=p'$.
The resulting $(\Delta,\xi)$ are thus labelled by 
a triple $\lbrace p (q,q') \rbrace$. 
In particular, we will consider below the case of
 $\Delta_1 =\Delta_{3 (1,1)}$ and $\xi_1=\xi_{3(1,1)}$.

The fusion rules are derived from applying the condition that
$\Phi_{\Delta_1, \xi_1}$ has a null descendant at level $\frac{3}{2}$. For ($\Delta_2,\xi_2$), we will consider a general member $\Phi_{p (q,q')}$.\footnote{Here we assume that $\Phi_{p (q,q')}$ has no extra null descendant other than those obtained from the nonrelativistic limit of the $(h_{p,q},{\bar{h}}_{p,q'})$ null states. While this is seen to be true for the level $\frac{3}{2}$, one needs to construct a formalism to verify this for any arbitrary level in the Kac table.} Thus we have from \refb{nreldelxi}, \refb{delta} and \refb{xi}:
\begin{align}
\Delta_1 &= \Delta_{3(1,1)} 
=  4 (1 - C_1)\,, \quad 
\xi_1 =\xi_{3(1,1)} 
= -4 C_2 \,;
\label{fus1} \\
\Delta_2 &= \Delta_{p(q,q^{\prime})}
= -{1 \over 2}C_1 \,(p^2-1) + {1 \over 8}\, [\,5 p^2 -2p(q+q') -1\,] \,,
\label{fus1-2} \\ 
\xi_2 &= \xi_{p(q, q^{\prime})}=-{1 \over 2} C_2\,(p^2-1)\,.
\label{fus1-3}
\end{align}

We need to consider the conditions \refb{nulleq1} and \refb{nulleq2} for the
case of the three point function. With 
\be{}
G^{(3)}_{\rm 2d\,SGCA}
(\{t_i,x_i,\alpha_i,\beta_i\})=\langle \,
\Phi_{\Delta_3,\xi_3}(t_3,x_3,\alpha_1,\beta_1)
\,\Phi_{\Delta_2,\xi_2}(t_2,x_2,\alpha_2,\beta_2) \,\Phi_{\Delta_1,\xi_1}(0,0,0,0)\,\rangle\,,\nonumber
\ee{}
these give the constraints:
\be{}
\Big[- \sum_{i=2}^{3} \Big\lbrace \frac{1}{t_{i}} (\alpha_i \p_{x_i}-\p_{\beta_i}) + \frac{2 \xi_i}{t^2_{i}}\,\alpha_i \Big\rbrace \,+\,\frac{1}{\xi_1} \sum_{i=2}^3\p_{x_i}\,\sum_{j=2}^3(\alpha_j \p_{x_j}-\p_{\beta_j}) \Big]\,G^{(3)}_{\rm 2d\,SGCA} = 0 \,, \nonumber
\ee{}
\ben{}
&& \Big[- \sum_{i=2}^{3} \Big\lbrace \frac{1}{t_{i}} (-\alpha_i \p_{t_i} -\beta_i \p_{x_i} +\,\p_{\alpha_i} )-\frac{\xi_i}{t^2_{i}}\,(-\alpha_i \p_{x_i} + \p_{\beta_i}) - \frac{2}{t_i^2}\,(\xi_i \beta_i-\Delta_i \alpha_i) + \, \frac{2\xi_i}{t_i^3}x_i \alpha_i\Big\rbrace \crcr
&& \,\,\,\,\, +\,\,\frac{1}{\xi_1} \sum_{i=2}^3\p_{x_i}\,\sum_{j=2}^3 (-\alpha_i \p_{t_i} -\beta_i \p_{x_i} +\,\p_{\alpha_i} ) - \,\frac{1}{\xi_1} \sum_{i=2}^3\p_{t_i}\,\sum_{j=2}^3 (\alpha_i \p_{x_j}-\p_{\beta_j} )\crcr
&&\,\,\,\,\,-\,\,\frac{\Delta_1 + 1}{\xi_1^2} \sum_{i=2}^3\p_{x_i}\,\sum_{j=2}^3(\alpha_j \p_{x_j}-\p_{\beta_j}) \Big]\,G^{(3)}_{\rm 2d\,SGCA} = 0 \,,\nonumber
\een{}
respectively.
Now by using \refb{nrel3pt}, these translate into
\ben{hogehoge} 
&&\xi_1\, (\xi_2+\xi_3) - (\xi_2- \xi_3)^2 =0\,, \crcr
&& (\Delta_2 + \Delta_3 -1)\, \xi_1^2- 2\,(\Delta_2-\Delta_3)\, (\xi_2 - \xi_3)\,\xi_1+ (\Delta_1 +1)\,(\xi_2 - \xi_3)^2 = 0\,. \nonumber 
\een

Solving the above equations, we get two simple sets of solutions:
\be{soln1}
\xi_3 =-{1 \over 2} C_2\,\lbrace(p\, \pm \, 2)^2-1 \rbrace\,,\, \Delta_3 =  -{1 \over 2}C_1 \,\lbrace (p\,\pm\,2)^2-1\rbrace + {1 \over 8}\, \lbrace 5 \,(p\,\pm \,2)^2 -2(p\,\pm\,2)\,(q\,+\,q') -1 \rbrace \,.
\ee{}
Comparing with \eq{fus1-2} and \eq{fus1-3}, we see that 
\be{deltwofus}
\Delta_3=\Delta_{p\,\pm \,2\,(q,q^{\prime})}\,, \qquad \xi_3=\xi_{p\,\pm\,2\,(q,q^{\prime})}\,,
\ee
which is exactly what the relativistic fusion rules imply, namely
\be{spfus}
[\Phi_{3(1,1)}] \times [\Phi_{p(q,q^{\prime})}]   =  [\Phi_{p+2\,(q,q^{\prime})}] +[\Phi_{p-2\,(q,q^{\prime})}]\,.
\ee{}
Thus once again we see evidence for the consistency of the SGCA limit of
the 2d SCFT. However, we would like to remind the reader that in the SCFT case, two independent fusion rules (dubbed ``even" and ``odd") arise (for each of the holomorphic and antiholomorphic sectors), as shown in \cite{sotkov}, and their composition gives the full fusion rule. This is due to the presence of two independent constants for the SCFT three point function in each sector. But we have seen in \refb{nrel3pt} that when we multiply the results for the two sectors and take the limit, the contributions coming from the Grassmann odd terms of the corresponding sectors do not survive. So in the context of SGCA, only the \textit{even} fusion rules of SCFT are relevant.

\section{Concluding Remarks}

This concludes our present study of the supersymmetric extension of the GCA in two dimensions. We found that 2d SGCFTs, with non-zero central charges $C_1$ and $C_2\,$, can be readily obtained by considering a somewhat unusual limit of a non-unitary 2d SCFT. While the resulting Hilbert space of the SGCFT  is again non-unitary, the theory seems to be otherwise well-defined. We found that many of the structures are parallel to those in the SuperVirasoro algebra and indeed arise from them when we realise the SGCA by means of the scaling limit. But in most cases we could also obtain many of the same results autonomously from the definition of the SGCA itself, showing that these are features of any realisation of this symmetry.

There are numerous avenues to explore in the study of nonrelativistic 2d theories, whose algebra can be obtained by group contraction of the well-studied relativistic theories. One of the immediate things one needs to understand better are the extra constraints arising purely within the nonrelativistic sector (as we have explained in Sec.~5.3), whose analogues do not not exist in the parent relativistic theory.\footnote{We would like to thank Rajesh Gopakumar and Ashoke Sen for valuable discussions on this point.} Our present understanding of the above issue is the following: This means that what was an irreducible representation (``irrep") of the (Super)Virasoro algebra (i.e. modulo the original null states) is no longer an irrep after taking the limit. This in itself is not surprising or unusual. The irreps of the original group need not to go over into irreps of the contracted group. It is therefore not too surprising that if we further choose to restrict to the irreps formed by modding out by the additional null states, there would be additional relations. The physical translation of these statements is that, if we choose to set the additional null states to zero, then the correlation functions have some further selection properties. Another way to put it is that the correlation functions of operators, which lie in the smaller vector space, may satisfy additional relations which would not be true of the full vector space. This is because we are choosing to work with a subclass of operators (states) which can close amongst themselves consistently rather than the full set of operators (states). The main thing to check is that the additional conditions are not incompatible. For all the specific cases we have dealt with in the present work, we have not found any inconsistency. Similar checks must be done for the higher levels, but at the moment we do not have a general way.

It is clear from the above discussion that it is not obvious to conclude that the (S)GCA arises as a limit of the (S)CFT without further analysis. We would like to stress that, in the present work as well as in \cite{2dgca}, we have not established in any strong way the existence of our limit of the (S)CFT. We have just performed a series of consistency checks. But one can look at the possibility whether one can construct a consistent (S)GCA where the extra relations do not play a role. The fusion rules found in GCA and SGCA indicate that one can truncate to the states in the usual Kac table. In other words, can the primaries (with the special values of $\xi$ and $\Delta$ corresponding to the extra null states) appear in the RHS of fusion rules of the other null state primaries? If they do not appear, then we think that we can consider a truncation where these kinds of null states do not have to be considered. We can then consider the family of primaries which have only the values in the nonrelativistic limit of the usual Kac table and the OPEs will close in this sector. We have found this to be true for the lowest level(s) where we get non-trivial null states. However, we have not proven this for states at any arbitrary level and we would like to explore whether it is possible to give a general proof that the fusion rules in the nonrelativistic theory always give other members of the original Kac table (and not anything else).

We would like to emphasize that we have not been able to provide any strong evidence of the presence of (S)GCA in possible field theories. This will also require proving our assumption of the state-operator correspondence.

The present work has been done focussing on the Neveu-Schwarz sector of the $N=(1,1)$ supersymmetric extension of GCFTs in 2d. One can try to work on the Ramond sector and find the analogous results there, where one cannot use the superfield formalism. Also, one can try to find out the consequences when we increase the number of supersymmetries. All these studies can be easily done along the framework presented in this work.

\subsection*{Acknowledgements}

I would like to thank Arjun Bagchi, Akitsugu Miwa, and especially Rajesh Gopakumar and Ashoke Sen for stimulating discussions. I would also like to express my gratitude to Rajesh Gopakumar and Ashoke Sen for their valuable comments on the manuscript and for patiently explaining me various intricate issues. This work was supported by the grant from Blaise Pascal Chair, France, of Professor Ashoke Sen.

\section*{Appendix}
\appendix

\section{Descendants and SGCA Conformal Blocks}

\label{Descendants}

\subsection{SGCA Descendants}

By means of the differential operators
$\hat M_{-n}$ , $\hat L_{-n}$ , $\hat{H}_{-r}$ and $\hat{G}_{-r}$ (with $m,n,r,s>0$) in  \refb{null-diff}, we may express the correlation function including a general SGCA descendant with the correlation function of the corresponding primary superfield
$\Phi_{\Delta\xi}(t,x,\alpha,\beta)$. We have in fact already used this in Sec.~6 as \refb{nulleq1} and \refb{nulleq2}, for the 
simple cases of these descendants corresponding to null states.
The general expression can be written as  
\ben{ope1}
&&\langle\,\Phi_k(t_k,x_k,\alpha_k,\beta_k)\,\cdots\,\Phi_2(t_2,x_2,\alpha_2,\beta_2)\,\Phi_1^{\lbrace  {\vec{l}}, {\vec{q}},{\vec{u}},{\vec{v}} \rbrace} (0,0,0,0) \,\rangle \crcr
&& \hspace{2mm} = {\hat L}_{-l_i} \cdots {\hat L}_{-l_1} 
\,{\hat M}_{-q_j}\cdots {\hat M}_{-q_1} \,  {\hat G}_{-u_{i'}} \cdots {\hat G}_{-u_1} \crcr
&& \hspace{9 mm} {\hat H}_{-v_{j'}}\cdots {\hat H}_{-v_1}  \, \langle\,\Phi_k(t_k,x_k,\alpha_k,\beta_k)\,\cdots\,\Phi_2(t_2,x_2,\alpha_2,\beta_2)\,\Phi_1 (0,0,0,0) \,\rangle \,,\crcr
&& \crcr
\mbox{for  }&& \Phi_1^{\lbrace  {\vec{l}}, {\vec{q}},{\vec{u}},{\vec{v}} \rbrace}(0,0,0,0) \,|0\rangle \crcr
&& ={ L}_{-l_i} \cdots { L}_{-l_1} 
\,{ M}_{-q_j}\cdots { M}_{-q_1} \,  { G}_{-u_{i'}} \cdots {G}_{-u_1}\, {H}_{-v_{j'}}\cdots {H}_{-v_1}\,\Phi_1(0,0,0,0) \,|0\rangle \,,\nonumber
\een{}
where 
\ben{}
&& {\vec{l}} = (l_1, l_2, \cdots ,l_i) \,,\quad{\vec{q}} = (q_1, q_2, \cdots ,q_j)\,, \crcr
&& {\vec{u}} = (u_1, u_2, \cdots ,u_{i'})\quad \mbox{and} \quad{\vec{v}} = (v_1, v_2, \cdots ,v_{j'})\nonumber
\een{}
are sequences of positive integers such that $l_1\leq l_2 \cdots \leq l_i$ and similarly for the $q$, $u$ and $v$'s. Also note that $\Phi_1^{\lbrace 0,0,0,0 \rbrace}(t_1,x_1,\alpha_1,\beta_1)$ denotes the primary $\Phi_1(t_1,x_1,\alpha_1,\beta_1)$ itself.

\subsection{The OPE and SGCA Blocks}

Just as in the relativistic case, the OPE of two SGCA primary superfields can be expressed in terms of the SGCA primary superfields and their descendants as
\be{ope}
\Phi_1(t,x,\alpha,\beta)\, \Phi_2(0,0,0,0) 
= \sum_{p}  \sum_{\lbrace {\vec{l}} , {\vec{q}},{\vec{u}},{\vec{v}} \rbrace}
C_{12}^{p\lbrace  {\vec{l}}, {\vec{q}},{\vec{u}},{\vec{v}} \rbrace}(t,x,\alpha,\beta)\,
\Phi_p^{\lbrace {\vec{l}}, {\vec{q}},{\vec{u}},{\vec{v}} \rbrace}(0,0,0,0) \, .
\ee
We should mention that, unlike in the case of a 2d SCFT, such an expansion is not analytic (see (\ref{GCAOPE}) below), as was also true for GCA in \cite{2dgca}.
The form of the two and three point functions clearly exhibit essential singularities. Nevertheless
we will go ahead with the expansion assuming it makes sense in individual segments such as $x, t>0$. 
One can find the first few coefficients $C_{12}^{p\lbrace  {\vec{k}},
{\vec{q}},{\vec{u}},{\vec{v}} \rbrace}(t,x,\alpha,\beta)$ by considering the three point 
function of the primary superfields $\langle \Phi_3 \Phi_1 \Phi_2 \rangle$. In such a situation one can replace $\Phi_1 \Phi_2$ in the three point function with the RHS of \refb{ope}, and obtain
\ben{opecor}
&&\langle \, \Phi_3(t',x',\alpha ', \beta ') \,\Phi_1(t,x,\alpha ,\beta)\,\Phi_2(0,0,0,0)\, \rangle  \crcr
&& =
\sum_{p, \lbrace {\vec{l}} , {\vec{q}},{\vec{u}},{\vec{v}} \rbrace}   
C_{12}^{p\lbrace {\vec{l}} , {\vec{q}},{\vec{u}},{\vec{v}} \rbrace}
(t,x,\alpha,\beta)\,\, \langle \,
\Phi_3(t',x',\alpha ',\beta ') \,
\Phi_p^{\lbrace  {\vec{l}}, {\vec{q}},{\vec{u}},{\vec{v}} \rbrace}(0,0,0,0) \,
\rangle \,. 
\een{}

We can find 
$C_{12}^{p\lbrace  0,0,0,0 \rbrace}$,
$C_{12}^{p\lbrace  0,0,1,0 \rbrace}$, $C_{12}^{p\lbrace  0,0,0,1 \rbrace}$,
$C_{12}^{p\lbrace 1,0,0,0 \rbrace}$, $C_{12}^{p\lbrace  0,1,0,0 \rbrace}$ and 
$C_{12}^{p\lbrace 0,0,1,1 \rbrace}$ by 
expanding the LHS of \refb{opecor} in powers of the parameter $t \over t'$ with $x' \over t'$, $x \over t$, $\frac {\alpha' \alpha}{t'}$, $\frac {\alpha' \beta}{t'}$ and $\frac {\alpha \beta '}{t'}$ as coefficients, and comparing the $\lbrace t',x',\alpha ',\beta '\rbrace$-dependence of both the sides. To make the final formulae simple, we concentrate on the case with
$\Delta_1=\Delta_2=\Delta$ and  $\xi_1=\xi_2=\xi $.

The expansion of the LHS is given as:
\ben{LHSexp}
&&\langle \,\Phi_3(t',x',\alpha ', \beta ')\, \Phi_1(t,x,\alpha ,\beta)\,\Phi_2(0,0,0,0)\,  \rangle \crcr
& =& C_{312} 
\, (t'-t-\alpha ' \alpha)^{-\Delta_3}\, t^{\,\Delta_3 -2 \Delta}\, {(-t')}^{-\Delta_3}
\exp \Big\lbrace \xi_3 \,{x' - x-\alpha ' \beta +\alpha \beta ' \over t'-t -\alpha ' \alpha}
+(2 \xi - \xi_3){x \over t}
+\xi_3 {x' \over t'} \Big\rbrace \crcr
&=& C_{312}'\, {t'}^{-2\D_3}\, {\rm e}^{2 \xi_3
{ x' \over t'}} \cdot  t^{\,\Delta_3 -2 \Delta}
\,{\rm e}^{(2 \xi - \xi_3) {x \over t}} \,
\textbf{\Big[} 1 + \Delta_3 {\alpha ' \alpha \over t'} + \xi_3 \,({\alpha \beta ' \over t'}-{\alpha ' \beta \over t'} +{\alpha ' \alpha \over t'}{x' \over t'}) \crcr
&&  \hspace{3mm} +\, \,\Big\{ \D_3 + \xi_3\,({x' \over {t'}} 
- {x \over t}) +\Delta_3 {\alpha ' \alpha \over t'} +\xi_3 \,({\alpha \beta ' \over t'}-{\alpha ' \beta \over t'} +2\,{\alpha ' \alpha \over t'}{x' \over t'} - {\alpha ' \alpha \over t'}{x \over t}) \Big\} \,{t \over t'}  \crcr
&& \hspace{3mm} +\, \Big\{ \Delta_3 {\alpha ' \alpha \over t'} + \xi_3 \,({\alpha \beta ' \over t'}-{\alpha ' \beta \over t'} +{\alpha ' \alpha \over t'}{x' \over t'}) \Big\} \,\Big\{ \D_3 + \xi_3\,({x' \over {t'}} - {x \over t})+\xi_3 \,({\alpha \beta ' \over t'}-{\alpha ' \beta \over t'})\Big\} \,{t \over t'} + {\cal O}((t/t')^2) \textbf{\Big]}\,,\crcr
&&
\een{}
where $C_{312}' = (-1)^{\Delta_3} \, C_{312}\,$.\\

The RHS is given by
\ben{RHS}
&&\sum_{p, \lbrace {\vec{l}} , {\vec{q}},{\vec{u}},{\vec{v}} \rbrace}   
C_{12}^{p\lbrace {\vec{l}} , {\vec{q}},{\vec{u}},{\vec{v}} \rbrace}
(t,x,\alpha,\beta)\,\, \langle \,
\Phi_3(t',x',\alpha ',\beta ') \,
\Phi_p^{\lbrace  {\vec{l}}, {\vec{q}},{\vec{u}},{\vec{v}} \rbrace}(0,0,0,0) \,
\rangle \crcr  
&=& \textbf{\Big[}\, C_{12}^{3\lbrace 0,0,0,0 \rbrace} (t,x,\alpha, \beta)\, +\,C_{12}^{3\lbrace 0,0,1,0 \rbrace} (t,x,\alpha, \beta)\, {\hat{G}}_{-\frac{1}{2}}\, +\, C_{12}^{3\lbrace 0,0,0,1 \rbrace} (t,x,\alpha,\beta)\, {\hat{H}}_{-\frac{1}{2}} \crcr
&& \quad +\,\, C_{12}^{3\lbrace 1,0 ,0,0\rbrace} (t,x,\alpha,\beta)\, {\hat{L}}_{-1} \,+\, C_{12}^{3\lbrace 0,1,0,0 \rbrace}(t,x,\alpha,\beta) \,{\hat{M}}_{-1} \crcr
&& \quad +\,\,C_{12}^{3\lbrace 0,0,1,1 \rbrace}(t,x,\alpha,\beta)\,{\hat{G}}_{-\frac{1}{2}}\,{\hat{H}}_{-\frac{1}{2}} + \ldots \textbf{\Big]}\, {t'}^{-2\D_3}\, {\rm e}^{2 \xi_3 {x' \over{t'}}}\crcr
&=& {t'}^{-2\D_3}\, {\rm e}^{2 \xi_3 {x' \over{t'}}} 
\textbf{\Big[} \,C_{12}^{3\lbrace 0,0,0,0\rbrace} \,+\,2 \,C_{12}^{3\lbrace 0,0,1,0\rbrace} (-\D_3 \alpha ' - \xi_3 {x' \alpha ' \over{t'}} + \xi_3 \beta ')\, {1 \over t'}-2 \xi_3 \,C_{12}^{3\lbrace 0,0,0,1\rbrace} \, {\alpha ' \over t'} \crcr
&& \hspace{3mm} +\, 2\,C_{12}^{3\lbrace 1,0,0,0\rbrace} ( \D_3 +  \xi_3 {x'\over{t'}}) {1 \over t'} + \,2 \xi_3\, C_{12}^{3\lbrace 0,1,0,0 \rbrace}\, {1 \over t'}+ 2\,\xi_3\, C_{12}^{3\lbrace 0,0,1,1\rbrace} (1 + 2\xi_3 {\alpha ' \beta ' \over{t'}} ) \,{1 \over t'} + \cdots \textbf{\Big]}\,.\nonumber \\
\een{}
One can easily read off the coefficients by comparing \refb{LHSexp} and \refb{RHS}\footnote{The reader should note that we have compared the full functional dependence on the coordinates $\lbrace t,x,\alpha,\beta, t', x',\alpha ' , \beta '\rbrace$ on both sides, though the LHS has been shown upto a certain order in $\frac{t}{t'}$ (which is just a convenient trick to extract out the expression for the $C_{12}^{p\lbrace  {\vec{l}},
{\vec{q}},{\vec{u}},{\vec{v}}\rbrace}$'s).}:
\ben{coef}
C_{12}^{3\lbrace 0,0,0,0 \rbrace} &=& C_{312}' \, t^{\,\Delta_3 -2 \Delta}\,
{\rm e}^{(2 \xi - \xi_3) {x \over t}} \,,\nonumber \\
C_{12}^{3\lbrace 0,0,1,0 \rbrace} &=& \frac{1}{2}\,C_{312}' \, t^{\,\Delta_3 -2 \Delta}\,
{\rm e}^{(2 \xi - \xi_3) {x \over t}}\, \alpha \,,\nonumber \\
C_{12}^{3\lbrace 0,0,0,1 \rbrace} &=& -\frac{1}{2} C_{312}' \, t^{\,\Delta_3 -2 \Delta}\,
{\rm e}^{(2 \xi - \xi_3) {x \over t}}\,\beta \,,\nonumber \\
C_{12}^{3\lbrace 1,0,0,0\rbrace} &=&  {1\over 2}\, C_{312}' \, t^{\,\Delta_3 -2 \Delta + 1}\, {\rm e}^{(2 \xi - \xi_3) {x \over t}}\,, \\
C_{12}^{3\lbrace 0,1,0,0 \rbrace} &=& - {1\over 2} \, C_{312}' \, x\,t^{\,\Delta_3 -2 \Delta}\, {\rm e}^{(2 \xi - \xi_3) {x \over t}} \,,\nonumber \\
C_{12}^{3\lbrace 0,0,1,1 \rbrace} &=& 0 \,.\nonumber
\een
So in this case, the SGCA OPE is
\ben{}
&&\Phi_1(t,x,\alpha,\beta)\, \Phi_2(0,0,0,0) \crcr
&=& \sum_{p} C_{p12} ' \, t^{\D_p - 2 \D} \, 
{\rm e}^{(2 \xi - \xi_p) {x \over t}} 
\Big( \Phi_p(0,0,0,0) + {\alpha \over 2} \,
\Phi_p^{\lbrace 0,0,1,0\rbrace}(0,0,0,0) - {\beta \over 2} \,
\Phi_p^{\lbrace 0,0,0,1\rbrace}(0,0,0,0) \crcr
&& \hspace{4.9 cm} + \, {t\over 2} \,
\Phi_p^{\lbrace {1}, {0} ,0,0\rbrace}(0,0,0,0) 
- {x\over 2}\, \Phi_p^{\lbrace  {0}, {1},0,0 \rbrace}(0,0,0,0) + 
\ldots \Big)\,. \nonumber \\ 
\label{GCAOPE}
\een{}


\end{document}